 \documentclass[final,5p,times,twocolumn,authoryear]{elsarticle}

\usepackage{amssymb}
\usepackage{lipsum}
\usepackage{amsmath}
\usepackage{float}
\usepackage{wasysym}
\usepackage{url}
\usepackage{xurl}
\usepackage{xcolor}
\usepackage{hyperref}
\usepackage{orcidlink}
\journal{High Energy Astrophysics}

\begin{document}

\begin{frontmatter}



\title{Cosmic Ray Diffusion and the Origin of  Very High Energy Gamma-Ray Emission in Young Massive Stellar Clusters}


\author[first]{Lucas Barreto-Mota\orcidlink{0000-0003-2164-9055}}
\affiliation[first]{organization={Instituto de Astronomia, Geofísica e Ciências Atmosféricas da USP},%Department and Organization
            addressline={Departamento de Astronomia, R. do Matão, 1226}, 
            city={São Paulo},
            postcode={05508-090}, 
            state={São Paulo},
            country={Brazil}}
\author[first]{Elisabete M. de Gouveia Dal Pino\orcidlink{0000-0001-8058-4752}}
% \affiliation[first]{organization={Instituto de Astronomia, Geofísica e Ciências Atmosféricas da USP},%Department and Organization
%             city={Earth},
\author[second]{Gaetano Di Marco\orcidlink{0009-0005-6018-2621}}
\affiliation[second]{organization={Instituto de Física Teórica UAM/CSIC, Departamento de Física Teórica, M-15, Universidad Autónoma de Madrid},
            addressline={Cantoblanco, 28049, Calle Nicolás Cabrera 13-15}, 
            city={Madrid},
            % state={Madrid},
            country={Spain}}

\author[third]{Rafael {Alves Batista}\orcidlink{0000-0003-2656-064X}}
\affiliation[third]{organization={Sorbonne Université, Institut d'Astrophysique de Paris},
            addressline={CNRS UMR 7095, 98 bis bd Arago 75014}, 
            city={Paris},
            % % state={},
            country={France}}

\author[first]{Stela Adduci Faria\orcidlink{0000-0002-6374-9321}}

\begin{abstract}
The search for Galactic sources capable of accelerating cosmic rays (CRs) to PeV energies has advanced significantly in recent years. In particular, high-energy observatories such as LHAASO have detected extended gamma-ray halos around several sources, suggesting that CRs may escape their acceleration sites through anomalously slow diffusion. Theoretical studies have proposed that magnetic mirror diffusion combined with pitch-angle scattering in the turbulent flow can naturally suppress CR transport. 
In this work, we first  discuss  how mirror diffusion combined with scattering can suppress cosmic-ray transport and naturally lead to an energy-dependent diffusion coefficient $D(E) \propto E^{1/3}$.
Then, we combine a 3D magnetohydrodynamic (MHD) simulation of a young massive stellar cluster (YMSC) with Monte Carlo CR propagation calculations (performed with CRPropa). The model includes the background gas density, magnetic field, stellar blackbody radiation and dust emission from the cluster, as well as the cosmic microwave background, and the Galactic interstellar radiation field.
Using the YMSC W43 as a benchmark, we compare two CR injection geometries: a central source and a spherical shell that represents the collective wind shock of the massive stars. We show that mirror+scattering diffusion $D(E) \propto E^{1/3}$, combined with a CR injection spectrum 
$E^{-2}$, reproduces the gamma-ray spectrum observed by Fermi and LHAASO. In contrast, stronger energy-dependent diffusion requires a harder CR injection spectrum, $\propto E^{ - 1.6}$, to match the data. We also find that the relative contributions of leptonic inverse Compton scattering and hadronic interactions depend sensitively on the assumed diffusion regime. Finally, the resulting spectra show no significant differences with regard to the injection CR source location, apart from the lower CR injection luminosity required in the central-source case. Overall, our  results indicate that the observed gamma-ray emission is shaped primarily by the diffusive propagation regime rather than by the precise location of the CR source. 

\end{abstract}




\end{frontmatter}

%\tableofcontents



\section{Introduction}
\label{sec:intro}

As Cosmic Rays (CRs) interact with the interstellar medium (ISM), they can leave a trace of high energy emission from the interactions with gas, photons and magnetic fields. Understanding the origin and production of this very-high-energy (VHE; $E \gtrsim 1 \; \text{TeV}$) emission is an important step in understanding the underlying physics that governs our universe. 

The recent detection by LHAASO of VHE emission at hundreds of TeVs is very intriguing, and compatible with the presence of several PeV sources of CRs (PeVatrons) inside our galaxy \citep{2021Natur.594...33C}. The potential sources linked to these observations include molecular cloud-supernova remnant (MC-SNR) interaction regions \citep[e.g.][]{PhysRevLett.108.051105,PhysRevLett.109.061101,2013Sci...339..807A}, pulsar wind nebula halos (also called PWN TeV halos) \citep{PhysRevD.104.123017,yan2023origin,2024NatAs.tmp...54Y}, and Young Star Clusters (YSCs) \citep{2021MNRAS.504.6096M,2021NatAs...5..465A}. 
The Cygnus cocoon, a very extended region surrounding a cluster of massive stars \citep{2021NatAs...5..465A}, is perhaps one of the most compelling example of this class. Its VHE emission likely originates from an expansive area surrounding the stars, likely due to diffusion away from them and into the ISM. 

Another striking source is the young massive stellar cluster (YMSC) W43, a \textit{mini-starburst} molecular cloud complex near sitting at around 5.5~kpc away from Earth. With a mass of \(\sim10^6 M_\odot\) and a bolometric luminosity of \(\sim10^7 L_\odot\), this region is among the most massive star-forming complexes in our Galaxy with the presence of several Wolf-Rayet and OB stars \citep{2010A&A...518L..90B,2016ApJ...828...32L,2022A&A...662A...8M,2024A&A...685A.101L}. Fermi-LAT has detected \(\gamma\)-ray photons in the GeV band, and more recently in the TeV range by LHAASO \citep{2025SCPMA..6879502C}, also making this source as a prime candidate as one the most powerful CR sources in our Galaxy.

One of the main challenges in interpreting these VHE emissions is accurately determining the CR diffusion coefficient. Observations by H.E.S.S. \citep{2018ApJ...866..143H} suggest that, in the vicinity of the Vela X pulsar wind nebula, CR electrons likely exhibit a diffusion coefficient of $\lesssim 10^{28}\,\mathrm{cm^2\,s^{-1}}$ at $10$~TeV within the inner region extending over a few tens of parsecs around the wind nebula. This value is at least two orders of magnitude smaller than the average Galactic diffusion coefficient typically inferred for the ISM at comparable energies \citep[e.g.][]{2007ARNPS..57..285S,2019PrPNP.10903710K}. HAWC observations of the Geminga and PSR B0656+14 pulsar wind nebulae point to a similarly suppressed diffusion regime \citep{2017Sci...358..911A}.

Strong assumptions are often invoked to explain these observations. These range from phenomenological parameterizations in which the diffusion coefficient is taken to be a fraction of the average ISM value estimated from quasi-linear theory \citep[QLT;][]{1966ApJ...146..480J,2002cra..book.....S}, to models based on self-confinement scenarios \citep[e.g.,][]{2002PhRvL..89B1102Y,2004ApJ...614..757Y,2007MNRAS.378..245B,2011ApJ...728...60B,2013ApJ...779..140X,2016A&A...588A..73C,10.1093/mnras/stz2089,2021ApJ...908..193M,2021ApJ...912..109K,2021MNRAS.502.5821F,2023arXiv230510251G,2023MNRAS.525.4985K,2023JPlPh..89e1701L,2023ApJ...952..168M,2024ApJ...961...80G}.

QLT faces well-known limitations in describing CR transport. In particular, it struggles to account for pitch-angle scattering near $\mu = 0$, corresponding to particles moving nearly perpendicular to the mean magnetic field, the so-called $90^\circ$ problem. As a result, standard QLT often underestimates the scattering efficiency and may overestimate the CR mean free path and diffusion coefficient. Earlier studies have explored several phenomenological turbulence models within the QLT framework \citep[see e.g.][]{1990JGR....9520673M,1999ApJ...520..204G,2016ApJ...830..130S}. However, both observations and numerical simulations indicate that these approaches may not provide sufficient scattering to reproduce the suppressed diffusion inferred around VHE gamma-ray sources \citep{2002ApJ...578L.117Q,Gabici_2019}.

Therefore, more realistic and thoroughly validated models of MHD turbulence are required. Such models can substantially modify pitch-angle scattering, acceleration, and spatial diffusion of CRs compared to earlier ad hoc prescriptions
\citep[e.g.][]{2002PhRvL..89B1102Y,2004ApJ...614..757Y,2007MNRAS.378..245B,2011ApJ...728...60B,2013ApJ...779..140X,2016A&A...588A..73C,2021ApJ...908..193M,2021ApJ...912..109K,2021MNRAS.502.5821F,2023MNRAS.525.4985K,2023JPlPh..89e1701L,2023ApJ...952..168M,2024ApJ...961...80G}.
Nevertheless, even QLT-based models and their extensions still face difficulties in reproducing the low diffusion coefficients required by observations
\citep[][]{2004ApJ...616..617S,2006MNRAS.373.1195L,2008ApJ...673..942Y,2009A&A...507..589S,2016A&A...588A..73C,2020MNRAS.493.2817K}.

The super-diffusive behavior of turbulent magnetic field lines \citep{1999ApJ...517..700L,2013ApJ...767L..39B,2013Natur.497..466E,2013ApJ...779..140X,2014ApJ...784...38L,2022MNRAS.512.2111H,2022ApJ...926...94M,2023ApJ...959L...8Z} and particle  interaction with magnetic mirrors \citep{1969ApJ...156..445K,2020ApJ...894...63X} have been combined in \cite{2021ApJ...923...53L} (henceforth LX21) to predict a new non-resonant diffusion mechanism which was termed ``mirror diffusion".

Mirror diffusion arises when compressible MHD turbulence generates magnetic mirrors that reflect CRs with large pitch angles ($\mu<\mu_c$, where $\mu_c$ is the angle cosine of the critical angle below which particles can interact with mirrors), causing them to diffuse slowly along the field lines rather than being trapped \citep{2021ApJ...923...53L}.
When mirroring and gyroresonant scattering act together, CRs alternate stochastically between slow mirror diffusion at large pitch angles and faster scattering diffusion at smaller ones, producing a L\'{e}vy-flight-like propagation that strongly enhances confinement \citep{2025ApJ...988..269B,2024ApJ...975...65Z,2025MNRAS.539.1236B}.

In this work we want to study how the properties of mirror diffusion in combination to pitch angle scattering may affect the VHE spectrum of  Galactic sources. For that we will model an YMSC
combining 3D MHD simulations and the Monte Carlo CR propagation and cascading. 

As noted above, YMSCs also require a suppressed diffusion coefficient to explain the observed $\gamma$-ray emission. Moreover, their role as PeVatrons remains under debate. In particular, the location and mechanism capable of accelerating particles to VHEs, as well as the environmental conditions inside these clusters, are still not well constrained. One possibility is that particles are accelerated by the collective winds of the massive stars of the cluster \citep[e.g.][]{2021MNRAS.504.6096M,2025A&A...695A.175M}.
Alternatively, the accelerated CRs may originate from a single compact object embedded in the center the cluster, such as a microquasar \citep{abaroa_microquasar_2026}. In either scenario, CRs must remain confined for sufficiently long times to reproduce the observed $\gamma$-ray spectrum,  and in this work we will investigate both scenarios, including the effects of mirroring.

This work is structured as follows, in Section \ref{sec:methodology} we discuss our simulations setup and considerations about the modeling of CR diffusion; in Section \ref{sec:results} we present the resulting SEDs obtained from our Monte Carlo simulations; in Section \ref{sec:discussion} we  compare our results with observations and previous results in the literature. Finally, in Section \ref{sec:conclusion} we present our conclusions.

\section{Numerical Method}\label{sec:methodology}
%%\label{}
% \lipsum[1]
\subsection{YMSC background MHD simulation}\label{sec:YMSC_sim_description}

\begin{figure*}
    \centering
    \includegraphics[width=1.8\columnwidth]{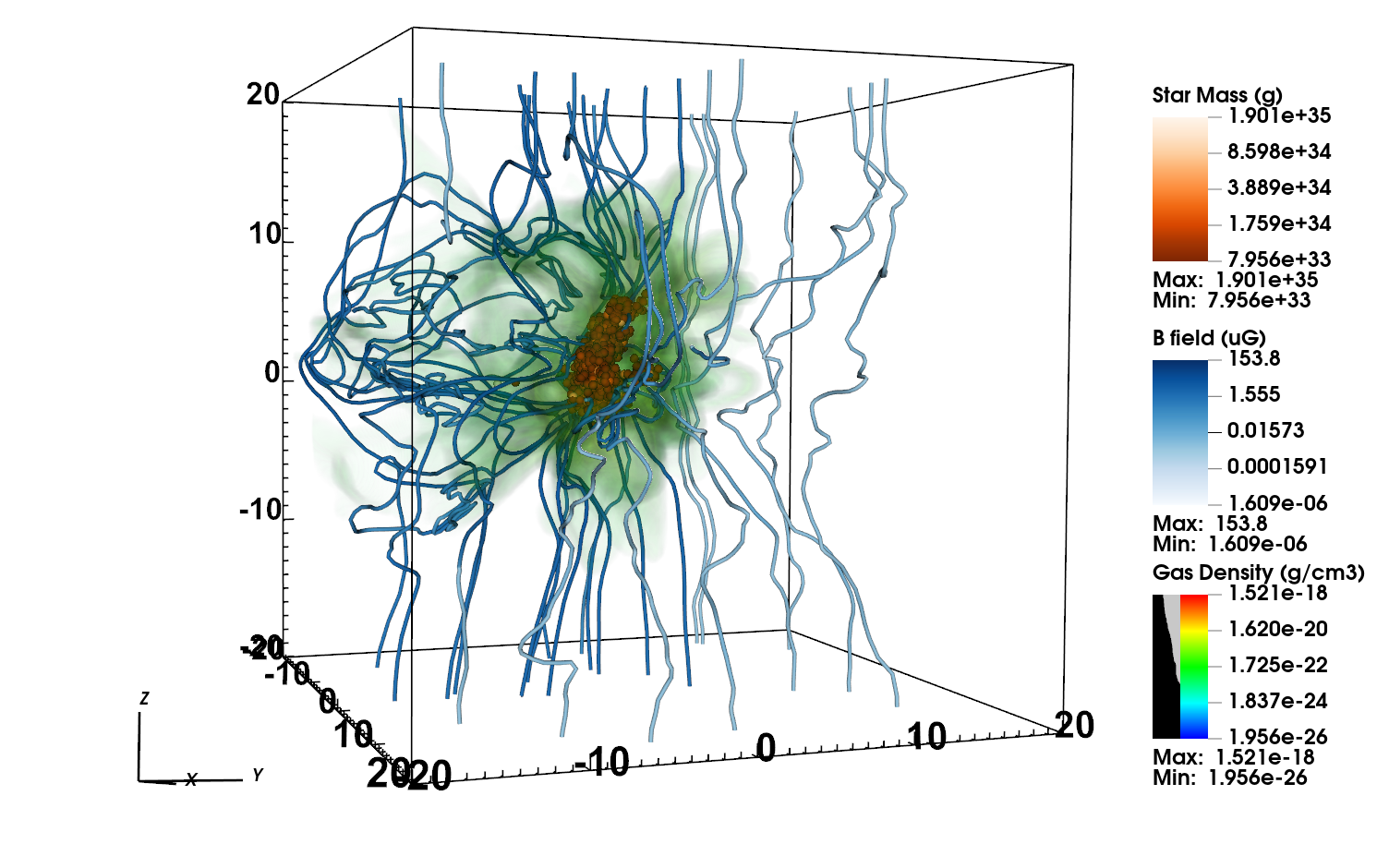}
    \caption{Background YMSCs of the simulation performed by \citet{2024A&A...690A..94P}. Red spheres show the position of the stars formed in the simulation. Green volumetric distribution shows the denser gas region. Blue and white lines show the magnetic field distribution. The axis scales are in pc.
    % The color grading is shown on the right-hand side of the image.
    }
    \label{fig:cluster_3D}
\end{figure*}

To model the background environment of a YMSC, we adopt the 3D MHD simulations of \citet{2024A&A...690A..94P}.
Their M5 simulation models a star cluster forming from an initially turbulent 
spherical gas cloud of mass $10^5\,M_\odot$ and radius 11.7~pc, with a central 
volume density $\rho_c = 28\,M_\odot\,\text{pc}^{-3}$, a free-fall time 
$t_\text{ff} = 2.1$~Myr, and an initial peak magnetic field $B_{0,z} = 1.85\,\mu$G. 
The simulation is run using the \textsc{Torch} framework, which couples MHD 
(\textsc{Flash}), $N$-body dynamics (\textsc{Petar}), and stellar evolution 
(\textsc{Seba}), and includes stellar feedback from radiation, winds, and supernovae. 
Stars are formed via sink particles sampling a Kroupa IMF in the mass range 
$0.08$--$100\,M_\odot$, with stars below $4\,M_\odot$ agglomerated into composite 
particles to reduce computational cost, resulting in over 15\,000 star particles 
tracked. The cluster reaches a final stellar mass of $6.5 \times 10^4\,M_\odot$ 
and is run to approximately $1.5\,t_\text{ff} \approx 3.1$~Myr, making it 
directly comparable to the W43 giant molecular cloud, which contains 
$1.32 \times 10^5\,M_\odot$ of gas within $R \sim 10$~pc \citep{2016ApJ...828...32L}, 
a cloud mass and radius closely matching those of the M5 initial conditions. Figure~\ref{fig:cluster_3D} shows  the gas density, the distribution of massive stars formed in the simulation, and the resulting turbulent magnetic-field structure of the nearly steady-state snapshot we considered for the Monte Carlo simulation  \citep[see ][for more details]{2024A&A...690A..94P}.

\subsection{CR Monte Carlo propagation}\label{sec:CRPropa-sims}

To study CR propagation and the associated $\gamma$-ray production inside the multi-zone,  YMSC background described in Section \ref{sec:YMSC_sim_description}, as well as the subsequent propagation of these $\gamma$-rays toward Earth, we use the three-dimensional Monte Carlo code CRPropa \citep{merten2017a,2022JCAP...09..035A}, extended with the hadronic interaction module presented in \cite{2025JCAP...04..043D}. CRPropa is a publicly available framework designed to model the propagation of very-high and ultra-high-energy CRs in both Galactic and extragalactic environments ({\UrlFont{https://crpropa.github.io/CRPropa3}}). It computes the lepto-hadronic cascading resulting from interactions of these CRs with the background environment (see Section \ref{subsec:background_photon_fields}). 

\subsection{Background Photon Fields}\label{subsec:background_photon_fields}

The MHD simulation of the YMSC shown in Figure~1 provides the background gas density and magnetic-field distributions through which CRs must propagate. In addition to these quantities, modeling CR propagation and interactions requires specifying the ambient photon fields within the cluster. These include the radiation from the massive stars and  dust emission in the YMSC, the Galactic interstellar radiation field, the cosmic microwave background, and Bremsstrahlung emission from the hot gas within the YMSC. In the following, we briefly describe each of these contributions.

\begin{figure}
    \centering
    \includegraphics[width=0.9\columnwidth]{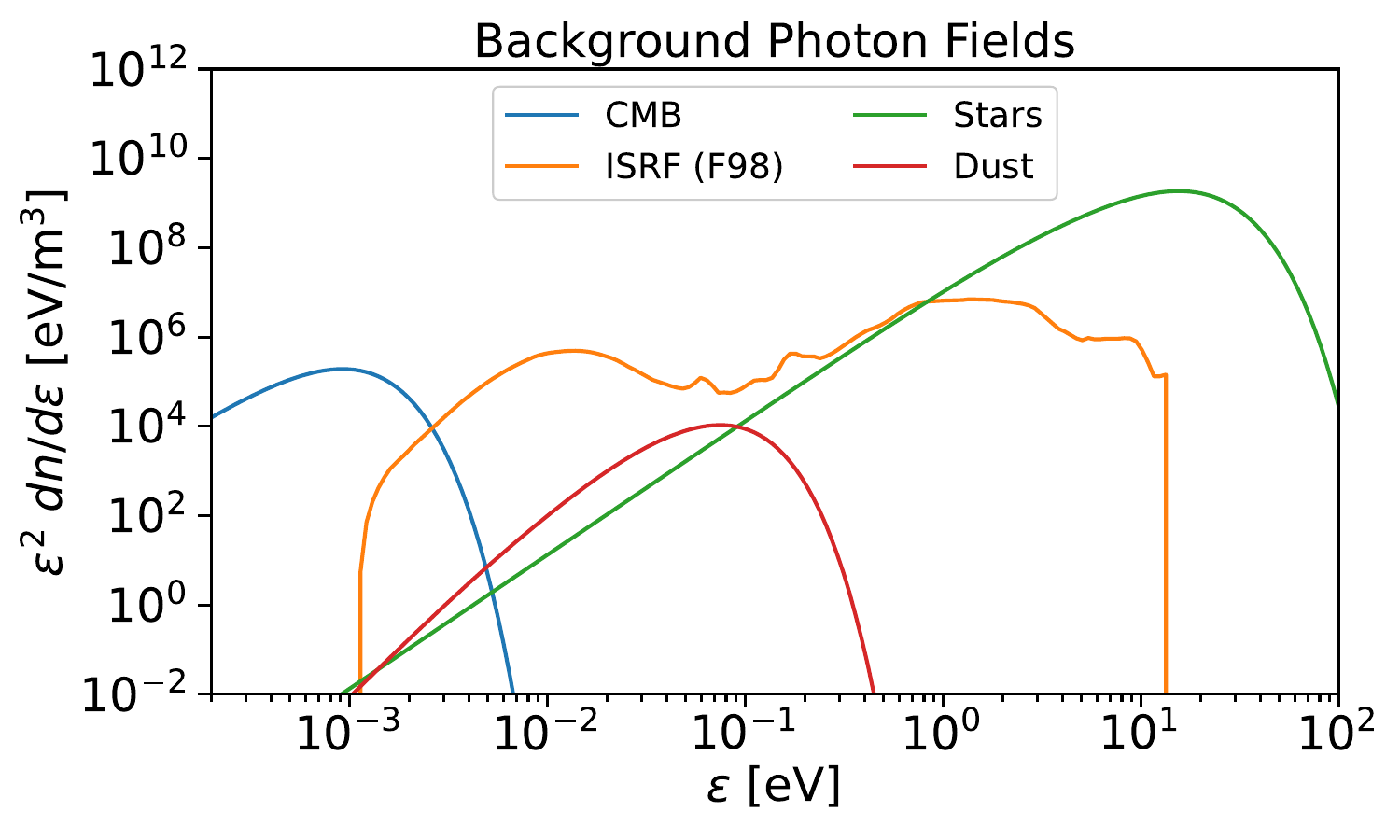}
    \caption{
    Energy density distribution of the background low-energy photon fields. The cosmic microwave background (CMB; blue) and the interstellar radiation (ISRF; orange) fields are already implemented the CRPropa code. The contribution from Bremsstrahlung radiation is several orders of magnitude smaller and, therefore, was not included in the plot. The photon field energy density calculation is described in the text and Appendix \ref{appendix:photon_field}.
    %[AUMENTAR AS LETRAS. E ONDE ESTÁ O BREMSSTRAHLUNG? SE NAO APARECE POR SER NEGLIGIBLE COMENTAR NO TEXTO.]}
    }
    \label{fig:all_photon_fields}
\end{figure}

%\bete{[\textbf{EXPLICAREMOS BREVEMENTE ABAIXO CADA UMA DAS COMPONENTES; E MOVA PARA O APENDICE O CALCULO EM DETALHE DE CADA PROCESSO RADIATIVO, DE MODO A REMOVER TODOS ESSAS SUBSECTIONS.}]}

\textit{Cosmic Microwave Background: }%\label{subsec:cmb} 
We include the cosmic microwave background (CMB) using the standard prescription implemented in CRPropa. The CMB is modeled as an isotropic blackbody photon field with temperature $T_{\rm CMB}=2.725\,{\rm K}$, whose spectral photon density follows the Planck distribution. This radiation field provides a homogeneous and well-characterized target for CR interactions and for $\gamma$-ray attenuation through processes such as pair production, depending on the particle species and energy range considered. 
We adopt the default CRPropa implementation of the CMB photon background \citep{2022JCAP...09..035A}, based on the standard blackbody description of the relic radiation field. 
%\citep{1996APh.....6...45P,2021APh...12602532N}.

\textit{Interstellar radiation field: }
The interstellar radiation field (ISRF) accounts for photons produced by stars and reprocessed by dust throughout the Galaxy. We model this component using the F98 model of \cite{2017ApJ...846...67P}, which combines optical and ultraviolet stellar emission with infrared dust emission to describe the large-scale Galactic radiation field. The ISRF constitutes an important target photon population for CR interactions, both through hadronic processes and inverse Compton scattering off CR electrons, as well as for \(\gamma\) attenuation via pair production at very high and ultra-high energies~\citep{2017ApJ...846...67P, PhysRevD.111.083004, aerdker2025crpropa}. Given the relatively short distance between W43 and Earth, however, the resulting pair-production opacity along this line of sight is small, and the spatial variation of the ISRF is not expected to significantly affect the propagated spectrum. We confirm this explicitly in Section \ref{sec:results}, where the inclusion of the propagation towards Earth is shown to have no appreciable impact on the simulated results.

\textit{Blackbody radiation from the stars inside the YSMC: }%
To model the radiation field produced by the stars in the YMSC, we assume that all stars present in the  3D MHD simulation emit as blackbodies. The details of the procedure are described in Appendix \ref{appendix:star_blackbody}.

\textit{Bremsstrahlung emission: }%\label{subsec:bremss}
The Bremsstrahlung contribution from%of
the hot gas inside the YSMC is described in Appendix \ref{appendix:bremsstrahlung_radiation}.  

\textit{Dust emission: }%\label{subsec:dust_emission}
The thermal dust emission from the YSMC is estimated assuming an isotropic optically thin distribution. The detailed calculation is presented in the Appendix \ref{appendix:dust_radiation}.

\textit{Extragalactic Background Light (EBL):}
Similar to the ISRF, the extragalactic background light (EBL) accounts for photons produced over the entire star formation history of the Universe, integrated from all extragalactic sources and reprocessed emission by dust. However, given its cosmological origin and consequent dilution over such large scales, its flux inside the cluster is many orders of magnitude smaller than the other photon field components, and therefore does not contribute appreciably to CR interactions within the cluster volume.

Figure \ref{fig:all_photon_fields} presents the corresponding photon energy density distributions as a function of the energy for each of these contributions. We note that the CMB dominates in the low-energy range, while radiation from the cluster stars and the Galactic ISRF dominates at the highest energies; the dust contribution is less relevant 
\footnote{We note that, if slightly different assumptions were adopted for the dust content in the YMSC, the peak of the emission could become comparable to that of the background ISRF. However, this would at most double the photon density around 0.1 eV and would not significantly affect the final $\gamma$-ray SED. This effect can be evaluated on a source-by-source basis and can be further explored in future work using more detailed descriptions of the photon fields.}. The photon density of the Bremsstrahlung component is several orders of magnitude below the other components and therefore was not included in Figure \ref{fig:all_photon_fields}.

\subsection{CR diffusion coefficient}\label{sec:cr_diffusion_coeff}

For simplicity, we parameterize the diffusion coefficient as
\begin{align}
    D(E) = D(E_0)
    \left( \frac{E}{E_0} \right)^\alpha ,
    \label{eq:diffusion_coefficient}
\end{align}
where  \(D(E_0)\) is the diffusion coefficient at a reference cosmic-ray energy
\(E_0\), and \(\alpha\) describes the energy dependence of cosmic-ray transport.
This dependence is determined by the dominant scattering and confinement
processes induced by the turbulent magnetic field.

As discussed in Section~\ref{sec:intro}, the functional form of \(D(E)\) depends
on the turbulence regime and on the magnetic-field geometry. 
In the 3D MHD model of the YMSC, the average Alfvénic Mach number is above 1, still the average magnetic field is of a few $\mu$G, with maximum above the hundreds of $\mu$G (see Figure \ref{fig:cluster_3D}). Moreover, the energy injected in the cluster provides a highly supersonic turbulence.
Under such conditions, fast MHD modes are expected to provide an efficient channel for cosmic-ray pitch-angle scattering. 
In addition, spatial variations in the magnetic-field strength
naturally generate magnetic mirrors, which can trap particles and further reduce
their effective diffusion. The physically preferred description for the present
environment is therefore one in which cosmic-ray transport is regulated by the
combined action of magnetic mirroring and pitch-angle scattering by fast MHD turbulence \citep{2021ApJ...923...53L}.
We adopt this mirror+scattering model as our reference case. In this
scenario, the diffusion coefficient follows a relatively weak energy dependence.
Following \cite{2021ApJ...923...53L}, who studied cosmic-ray transport in a mirror+scattering regime, we write
\begin{align}
    D(E) = D_{1{\rm PeV}}
    \left( \frac{E}{1\,{\rm PeV}} \right)^{1/3} ,
    \label{eq:diffusion_coefficient_mirror+scatter}
\end{align}
where we adopt \(D_{1{\rm PeV}}=10^{29}\,{\rm cm^2\,s^{-1}}\) from \citealt{2021ApJ...923...53L} \citep[see also][]{2025ApJ...988..269B}. Therefore, our fiducial model assumes \(\alpha=1/3\).
Magnetic mirroring increases the residence time of the particles, while pitch-angle scattering prevents free streaming along the magnetic field. Together, these effects provide strong confinement and make this prescription the most appropriate one for the YMSC conditions considered here\footnote{We note that \citealt{2025ApJ...988..269B}, adopting a mirror+scattering diffusion scenario, derived values of \(D_{1{\rm PeV}}\sim {\rm a~few}\times 10^{27}\,{\rm cm^2\,s^{-1}}\) from 3D MHD numerical simulations of sub-Alfv\'enic turbulence applied to pulsar wind nebulae. These values are particularly consistent with those required by observations of such systems \citep{2022FrASS...922100F}. In this work we  use the expected coefficients from \citep{2021ApJ...923...53L}, but the diffusion can be slower depending on the environmental conditions.}.
Notably, this dependence is the same as that usually assumed in models where CRs diffuse through a Kolmogorov-like, compressible turbulence \citep[e.g.][]{2025icrc.confE.692I}.

For completeness, and to assess the sensitivity of our results to the assumed
transport prescription, we also consider two alternative diffusion models. These
are included only as comparison cases, rather than as equally favored
descriptions of the present system:
\begin{itemize}
    \item \(\alpha=0.7\), \(D_{1{\rm PeV}}\simeq 10^{28}\,{\rm cm^2\,s^{-1}}\): this case represents a regime in which magnetic
    mirroring dominates the transport, without efficient pitch-angle scattering, 
   \begin{align}
   D(E) = D_{1{\rm PeV}}
   \left( \frac{E}{1\,{\rm PeV}} \right)^{0.7} ,
   \label{eq:diffusion_coefficient_mirror_only}
\end{align}
where  \(D_{1{\rm PeV}}\simeq 10^{28}\,{\rm cm^2\,s^{-1}}\) 
\citep{2021ApJ...923...53L, 2025ApJ...988..269B}.
     We regard this scenario
    as less favored than the mirror+scattering model because fast-mode
    turbulence is expected to contribute to scattering in the YMSC environment.
    It is nevertheless useful as an intermediate comparison with other diffusion
    prescriptions adopted in the literature.
    \item \(\alpha=1.0\), \(D_{1{\rm PeV}}\simeq 3.3\cdot10^{28}\,{\rm cm^2\,s^{-1}}\): this corresponds to a Bohm-like scaling \citep[e.g.][]{longair2011high} and is
    used only as a limiting comparison case:
     \begin{align}
    D(E) = D_{1{\rm PeV}}
    \left( \frac{E}{1\,{\rm PeV}} \right) ,
    \label{eq:diffusion_coefficient_bohm}
\end{align}
where  \(D_{1{\rm PeV}}\simeq 3.3\cdot10^{28}\,{\rm cm^2\,s^{-1}}\).% (ref.??).
 
 Although Bohm diffusion is often
    adopted as an extreme prescription for strong scattering, it does not
    specifically capture the combined mirror trapping and fast-mode scattering
    expected in the MHD turbulent model.
\end{itemize}
Thus, throughout this work, the \(\alpha=1/3\) mirror+scattering case is
treated as the physically preferred model for cosmic-ray diffusion in the YMSC\footnote{We note that other effects are likely present in such systems, but given our assumptions they would be secondary when compared to the combination of mirror diffusion with pitch angle scattering. See for instance Appendix B in \cite{2025ApJ...988..269B}. See also complementary studies of turbulence driven CR diffusion such as 
\cite{2024ApJ...975...65Z}, \cite{2023MNRAS.525.4985K}, and \cite{2023JPlPh..89e1701L}.}, whereas the \(\alpha=0.7\) and \(\alpha=1.0\) cases are included only to illustrate how the results vary under less favored transport assumptions.

\subsection{Location of the CR source inside the cluster}
We aim to investigate how diffusion affects the final $\gamma$-ray spectrum observed at Earth. To this end, we consider two scenarios for the CR source location: a point-like source placed at the center of the YMSC, and a spherical shell with a radius of 10 pc. This radius encloses most of the massive stars formed in the cluster and is compatible with a scenario in which CRs are produced at the shock front of the collective stellar wind \citep[see e.g.][]{2021MNRAS.504.6096M,2025A&A...695A.175M}.

In the central-source scenario, particle acceleration is driven by a compact object embedded in the cluster, such as a microquasar or a pulsar \citep{abaroa_microquasar_2026}. In this case, CRs may be accelerated by mechanisms such as shock acceleration \citep[e.g.][]{1977ICRC...11..132A,1978MNRAS.182..147B} or magnetic reconnection 
\citep[e.g.,][]{2005A&A...441..845D,2011ApJ...735..102K, 2012PhRvL.108x1102K,2024arXiv241013071D}.

In both scenarios, we assume that the CRs are injected with a power-law spectrum: 

\begin{align}\label{eq:injection_dist}
    \frac{dN}{dE}\Bigg|_{\text{inj}} \propto E^{-\alpha_{CR}} \cdot \exp\left(-\frac{E}{E_{\text{cutoff}}}\right),
\end{align}
where $\alpha_\text{CR}$ is the spectral index and $E_{\text{cutoff}}$ the cutoff energy of the CRs.  

This CR spectrum is injected with a power $P_{CR}$ which is a given fraction, $\eta_{CR}$, of the total bolometric luminosity of the YMSC.  This fraction  together with $\alpha_{CR}$ and $E_{\text{cutoff}}$, are free parameters of our model and the  adopted values are discussed in Section \ref{subsec:initial-cond} below. 

\begin{figure}
    \centering
    \includegraphics[width=1.0\columnwidth]{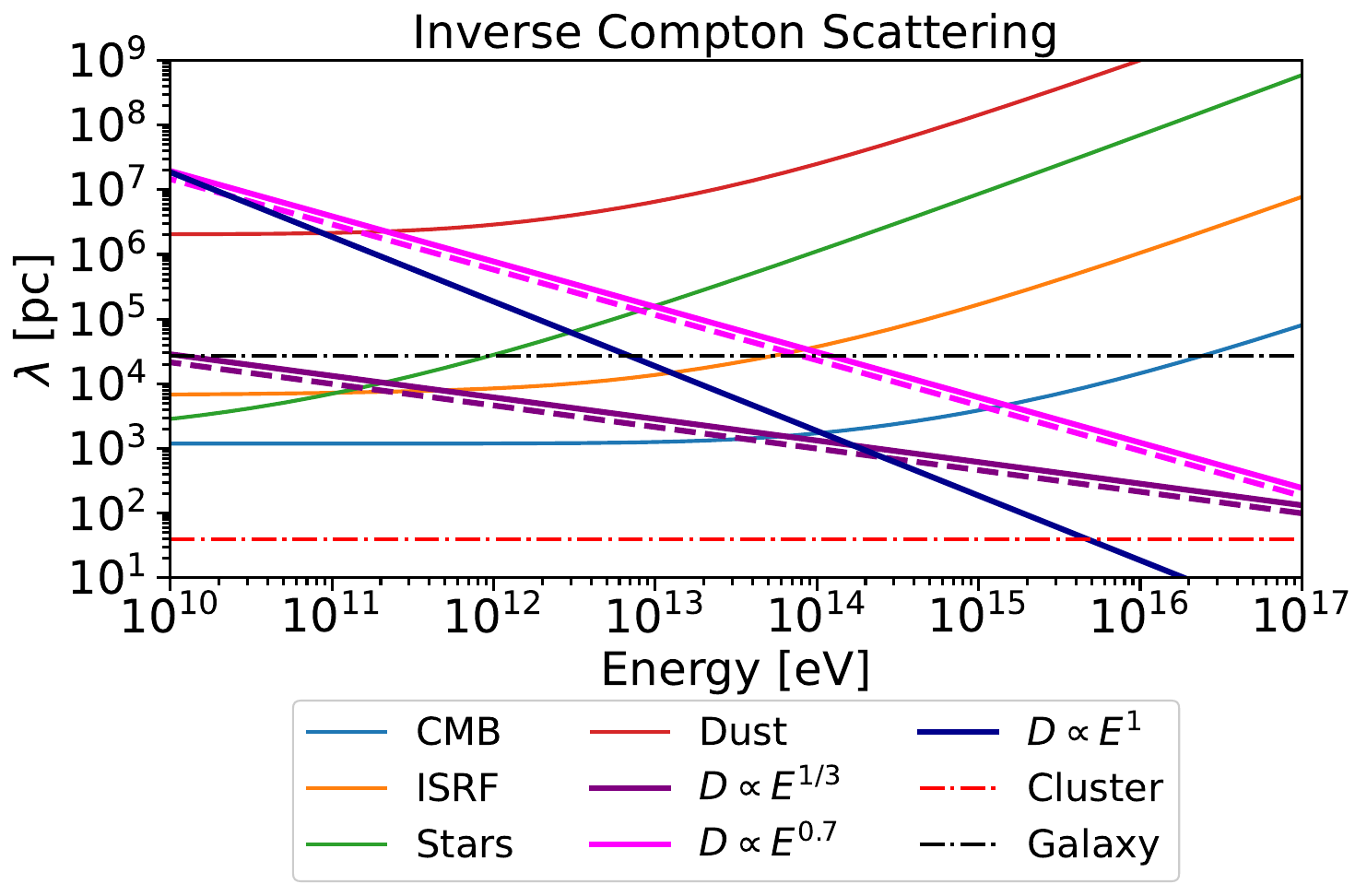}
    \caption{Mean free path for inverse Compton scattering, for the different background photon fields, compared to the effective particle propagation length inside the stellar cluster, $L_{\rm eff}$, both as functions of the particle energy. Blue line shows the mean free path for interactions with the CMB, orange for the ISRF, green line for the photon field from the stellar cluster YMSC, and red solid line for the dust photon field. 
    Dash-dotted red and black lines show the physical size of the simulated cluster and  of the  Galaxy, respectively. Purple, magenta, and dark blue lines show the effective propagation lengths, $L_{\rm eff}$, for the modeled diffusion coefficients.  
    Solid lines show $L_{\rm eff}$ for the CR central source, and dashed ones for the shell-like source. See text for further details. See also Appendix \ref{appendix:sync_loss} for a discussion on the synchrotron losses.
    }
    \label{fig:EMIC_intrate}
\end{figure}

\begin{figure}
    \centering
    \includegraphics[width=1.0\columnwidth]{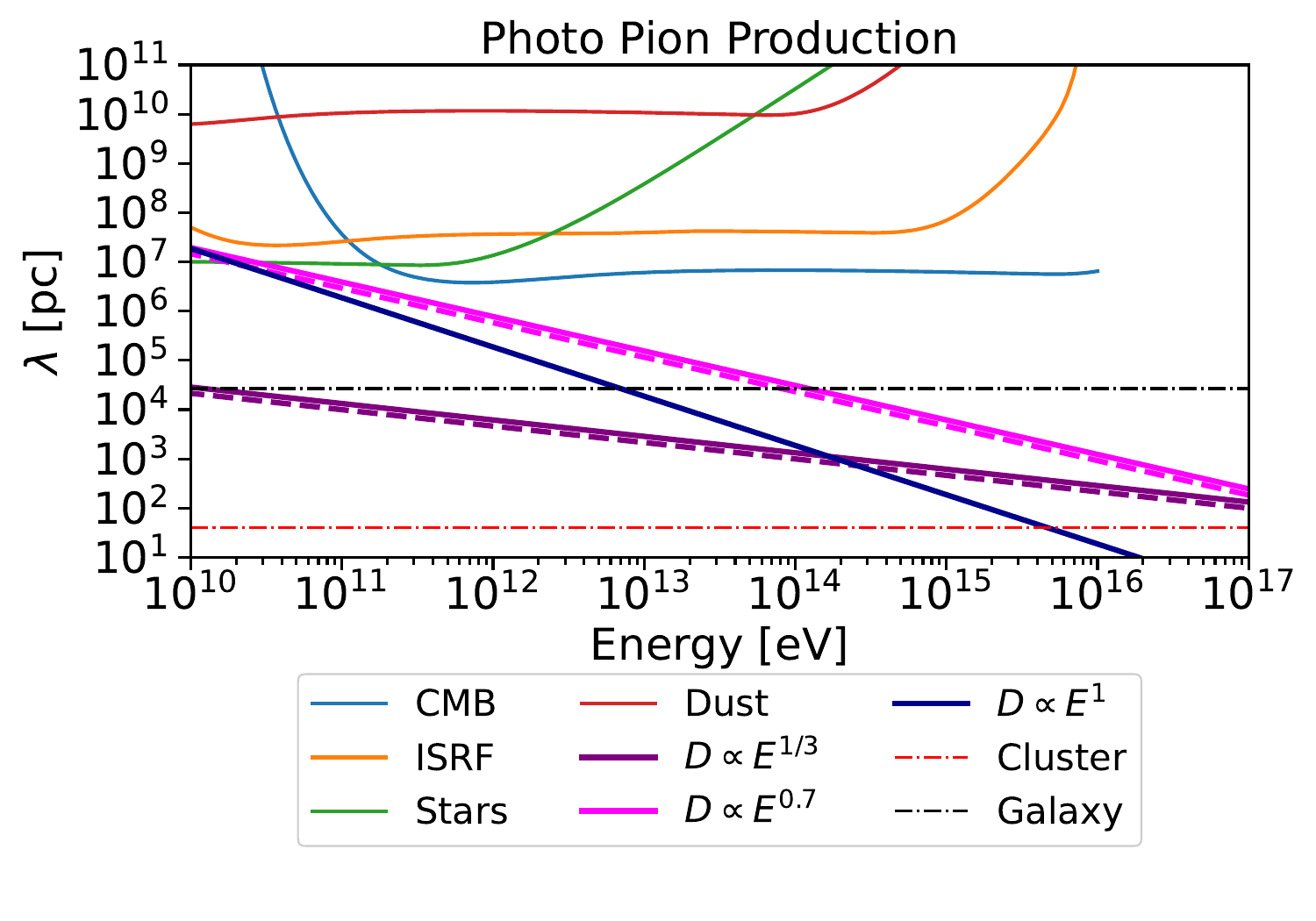}
    \caption{Mean free path for photo-pion production. The color scheme is the same as in Figure \ref{fig:EMIC_intrate}. }
    \label{fig:EM_photopion_prod}
\end{figure}

\begin{figure}
    \centering
    \includegraphics[width=1.0\columnwidth]{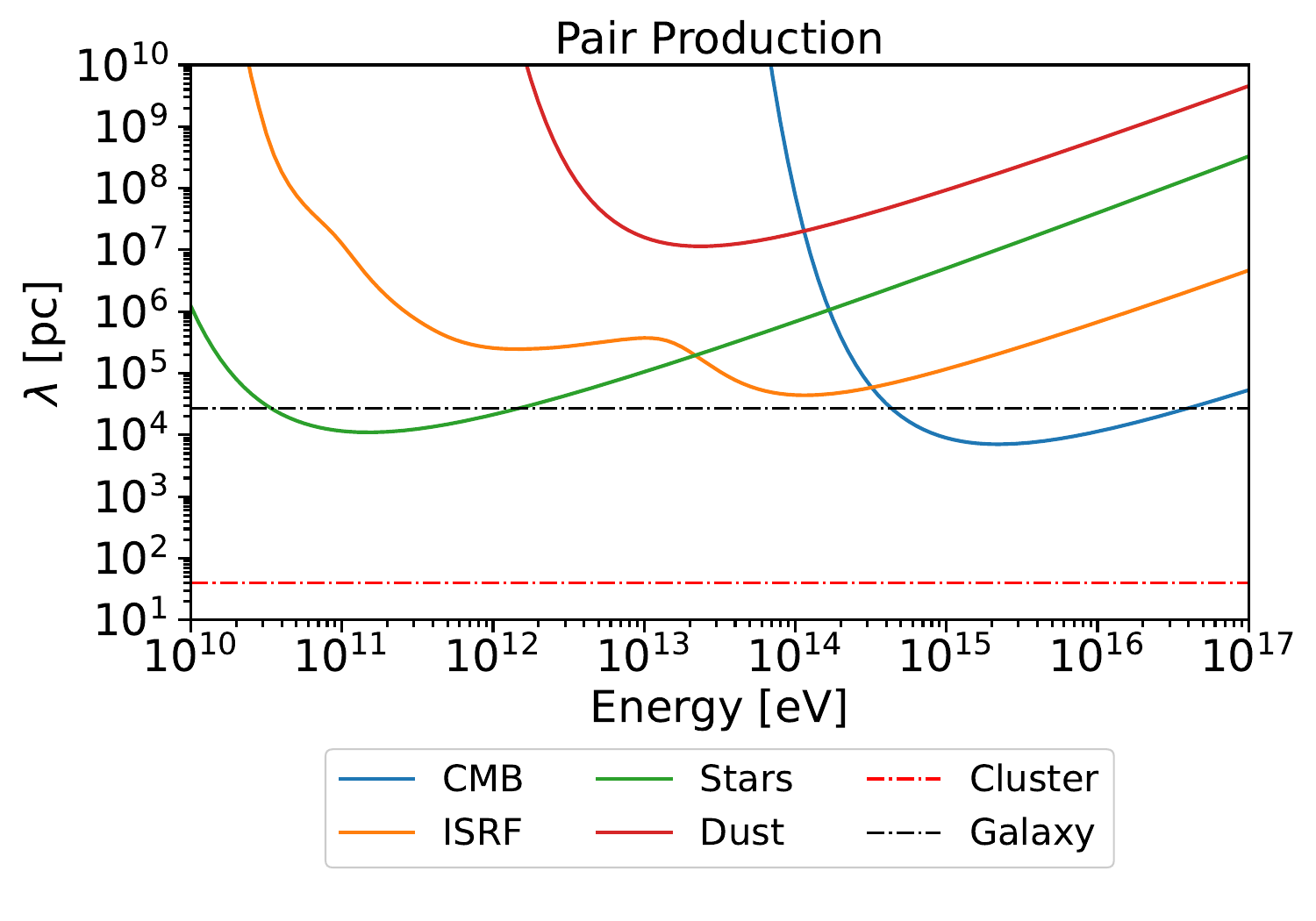}
    \caption{Mean free path for electromagnetic pair production. The color scheme is the same as in Figure \ref{fig:EMIC_intrate}.}
    \label{fig:EM_pair_prod}
\end{figure}

\begin{figure}
    \centering
    \includegraphics[width=1.0\columnwidth]{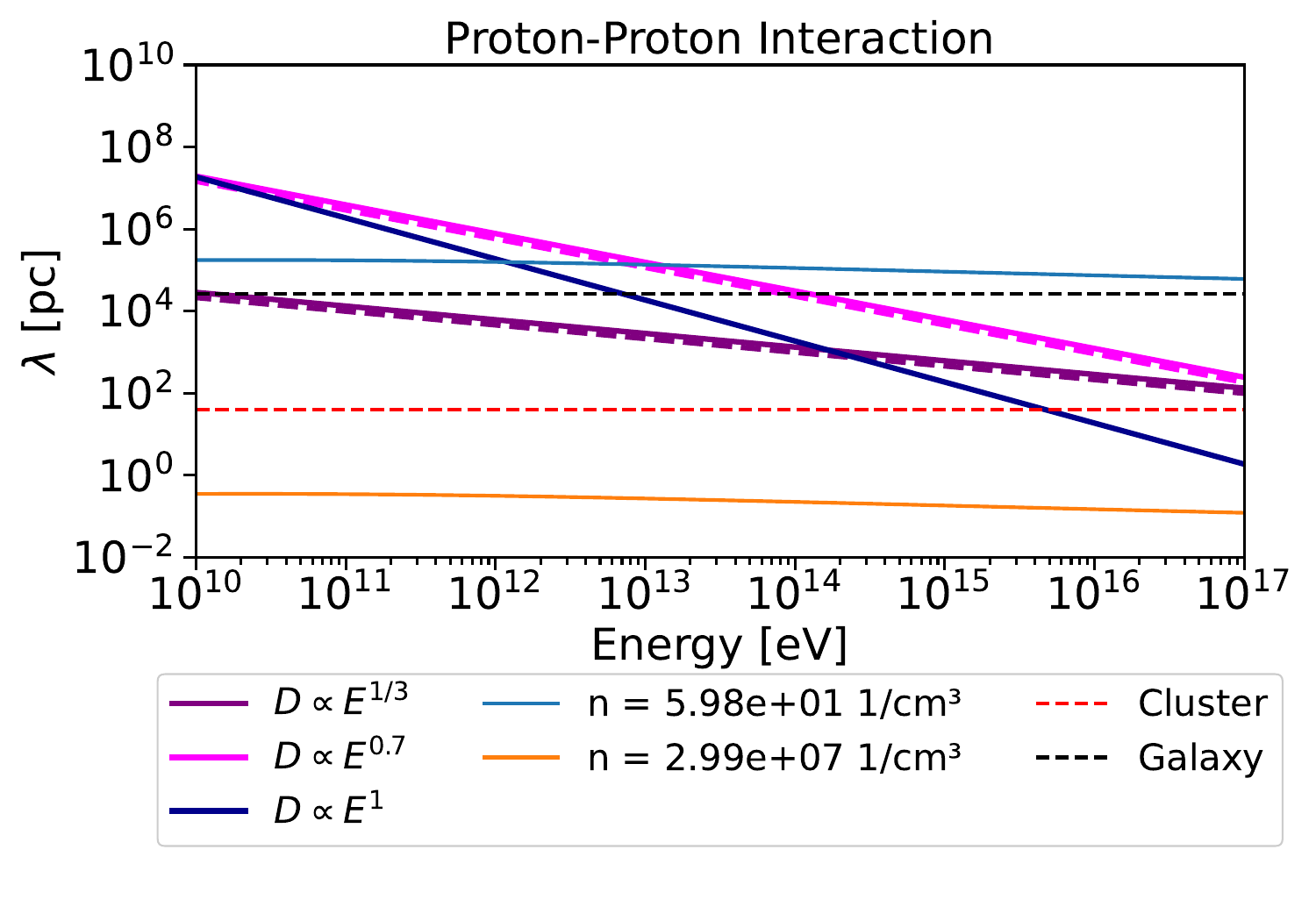}
    \caption{Mean free path for proton-proton interaction considering the lowest (light-blue line) and highest density (orange) values present in the YMSC simulation considering the cross-section of \cite{2006PhRvD..74c4018K}. Dashed red and black lines show the size of our simulated cluster and the size of our Galaxy, respectively. Purple and magenta lines show the effective size of the cluster for a particle with a given energy, depending on the modeled diffusion coefficient. Solid lines show the effective size for a central source, dashed ones show the effective size for the shell source. See text for further details 
    }
    \label{fig:PP_intrate}
\end{figure}

\section{Interaction rates of the relevant non-thermal processes}
\label{sec:interaction_rates}
In order to identify the dominant radiative and hadronic channels responsible
for the non-thermal emission from the YMSC, we first compare the interaction
mean free paths of cosmic rays and secondary photons with the characteristic
propagation scale inside the cluster. This comparison allows us to determine
which processes are efficient before the particles escape from the source
region.
For a cosmic-ray particle of energy \(E\) propagating through an isotropic
background photon field with differential number density \(dn/d\epsilon\), the
inverse interaction mean free path can be written as
\begin{equation}
\lambda^{-1}(E,z)
=
\frac{1}{8 \beta E^2}
\int\limits_0^\infty d\epsilon
\int\limits_{s_{\rm min}}^{s_{\rm max}} ds\,
\frac{1}{\epsilon^2}
\frac{dn(\epsilon,z)}{d\epsilon}
\,F(s),
\label{eq:mfp_general}
\end{equation}
where $z$ is the redshift of the source, 
\(\epsilon\) is the background photon energy, \(s\) is the squared
center-of-mass energy for a particle of mass $m$ moving with velocity $\beta c$, and \(F(s) = (s - m^2c^4) \sigma(s) \) contains the relevant cross section ($\sigma$).
As discussed in Section \ref{subsec:background_photon_fields} (Figure \ref{fig:all_photon_fields}), in this  work, the relevant photon
fields are the CMB, the ISRF, and the radiation field produced by the massive stars in the cluster.
The main photon-induced processes considered here are:
\begin{align*}
    \gamma + \gamma_{\rm bg} &\rightarrow e^{+}+e^{-},
    &&\text{pair production}, \\
    e^{\pm}+\gamma_{\rm bg} &\rightarrow e^{\pm}+\gamma,
    &&\text{inverse Compton scattering}, \\
    p+\gamma_{\rm bg} &\rightarrow 
    \begin{cases}
        p + \pi^{0},\\
        n + \pi^{+},
    \end{cases} &&\text{photo-pion production}
\end{align*}
The neutral pions decay into gamma rays,
\begin{equation*}
    \pi^0 \rightarrow 2\gamma,
\end{equation*}
whereas charged pions produce secondary leptons and neutrinos through
\begin{equation*}
    \pi^{\pm} \rightarrow \mu^{\pm}+\nu_{\mu}(\bar{\nu}_{\mu}),
\end{equation*}
followed by
\begin{equation*}
    \mu^{\pm} \rightarrow e^{\pm}
    + \nu_e(\bar{\nu}_e)
    + \bar{\nu}_{\mu}(\nu_{\mu}) .
\end{equation*}
These secondary electrons and positrons can subsequently contribute to the
gamma-ray emission through inverse Compton scattering.

In addition to photon-induced interactions, we also include inelastic
proton--proton collisions,
\begin{equation*}
    p+p \rightarrow p+p+\pi^{0,\pm}+X,
\end{equation*}
where $X$ denotes additional hadronic byproducts (e.g., kaons and higher-mass resonances). which are particularly important in dense environments such as YMSCs. This channel is particularly important in dense environments such as YMSCs. It produces gamma rays through \(\pi^0\) decay and secondary leptons through charged-pion decay. Therefore, proton--proton interactions can contribute both directly to the gamma-ray spectrum and indirectly through secondary inverse Compton emission.

To assess whether a given interaction is efficient inside the cluster, we
compare its mean free path, \(\lambda(E)\), with the effective propagation
length of particles inside the source. Because the particles diffuse rather
than propagate ballistically, the relevant path length can be much larger than
the physical size of the cluster. For a source at the center of a cluster of
radius \(R\), the characteristic diffusive escape time is
\begin{equation}
    t_{\rm esc}(E) \simeq \frac{R^2}{6D(E)} ,
\end{equation}
so that the effective propagation length is
\begin{equation}
    L_{\rm eff}(E) \simeq c\,t_{\rm esc}(E)
    = \frac{cR^2}{6D(E)} .
    \label{eq:Leff_central}
\end{equation}
For a shell-like source located at radius \(R_s\), the relevant distance to
escape is approximately \(R-R_s\), giving
\begin{equation}
    L_{\rm eff}(E) \simeq
    \frac{c(R^2-R_s^2)}{6D(E)} .
    \label{eq:Leff_shell}
\end{equation}
Thus, slower diffusion corresponds to a larger effective propagation length
and therefore to a higher probability of interaction before escape.

Figures~\ref{fig:EMIC_intrate}, \ref{fig:EM_photopion_prod}, \ref{fig:EM_pair_prod}, and
\ref{fig:PP_intrate} compare the interaction mean free paths with the physical
size of the cluster, the size of the Galaxy, and the effective diffusive
propagation lengths associated with the diffusion models discussed in this work.
The horizontal red dashed line marks the physical size of the simulated YMSC,
while the black dashed line indicates a Galactic scale. The purple, magenta,
and dark-blue curves represent \(L_{\rm eff}(E)\) for the diffusion
coefficients \(D(E)\propto E^{1/3}\), \(D(E)\propto E^{0.7}\), and
\(D(E)\propto E\), respectively. 
Solid curves correspond to a central CR source,
whereas dashed curves correspond to a shell-like source. We have also evaluated the synchrotron loss scales for the different diffusion regimes, see Appendix \ref{appendix:sync_loss} for more details.

Figure~\ref{fig:EMIC_intrate} shows the mean free path for inverse Compton
scattering. 
The comparison between \(\lambda_{\rm IC}\) and
\(L_{\rm eff}\) indicates that inverse Compton losses can be relevant over a
broad range of particle energies, especially when diffusion is 
slower.
The stellar photon field of the YMSC is particularly important at smaller energies because of its large photon energy density inside the cluster consistent with Figure \ref{fig:all_photon_fields}. 
CMB and ISRF also contribute, but for broader energy intervals, reflecting
their different characteristic photon energies. 
Therefore, inverse Compton
scattering is expected to provide a non-negligible contribution to the final
gamma-ray spectrum in all diffusion scenarios considered here.

Figure~\ref{fig:EM_photopion_prod} presents the corresponding mean free path for
photo-pion production. This process is generally much less efficient than inverse
Compton scattering inside the cluster, because the interaction mean free paths
are typically larger than both the physical size of the source and the effective propagation lengths. 

Figure~\ref{fig:EM_pair_prod} shows the mean free path for electromagnetic pair
production, \(\gamma\gamma\rightarrow e^+e^-\). In this case, the interaction
mean free paths are generally larger than the cluster scale for the
energy range relevant to the observed gamma-ray emission. This indicates that
 gamma-ray absorption inside the YMSC is weak. Pair production may
become relevant  during propagation through
larger Galactic scales, but it does not strongly suppress the gamma-ray flux produced inside the cluster itself.

Finally, Figure~\ref{fig:PP_intrate} shows the mean free path for proton--proton
interactions. Since this process depends on the gas density, we show two
limiting cases corresponding to the lowest and highest density values present
in the YMSC simulation. The large contrast between these curves reflects the
strong spatial inhomogeneity of the gas distribution. In the densest regions,
the proton--proton mean free path becomes much shorter than the effective
diffusive propagation length for all models, indicating that hadronic interactions are highly
efficient. This makes proton--proton collisions the dominant channel for
secondary particle production in dense parts of the cluster. The resulting
neutral-pion decay provides a direct gamma-ray component, while charged-pion
decay produces secondary electrons and positrons that can further radiate
through inverse Compton scattering.

Overall, Figures~\ref{fig:EMIC_intrate}--\ref{fig:PP_intrate} show that the relative
importance of the different non-thermal processes is controlled by the
competition between interaction mean free paths and diffusion-regulated
residence times. Inverse Compton scattering is relevant in broad energy range for all models, 
photo-pion interactions are not expected to contribute,
and internal pair production is not efficient on the cluster
scale. The dominant production of secondaries, 
arises from proton--proton interactions in the dense gas of the YMSC. This result
motivates the detailed gamma-ray spectral calculations presented in the next section.

As a final remark, we note that in this work we assume isotropic photon fields arising from the cumulative stellar emission of the cluster. While each individual star emits radiation approximately isotropically, the photon field produced by a given star is anisotropic at any location within the cluster, since photons propagate along the line connecting the star and the interaction site. Moreover, the thermal emission from dust depends on the local dust distribution and is therefore also expected to exhibit spatial variations. Consequently, a more realistic treatment would require Equation~\ref{eq:mfp_general} to depend both on the position within the cluster and on the interaction angle between the cosmic ray (or $\gamma$-ray) and the background photon field, leading to position- and angle-dependent interaction rates. Nevertheless, because we are interested in the cumulative radiation field produced by the ensemble of cluster stars, the isotropic approximation provides a reasonable  description, particularly for the large-scale properties of the system. Although incorporating the full angular and spatial dependence of the radiation field is beyond the scope of the present work and is left for future investigation, the isotropic approximation captures the essential physics relevant to our calculations, as demonstrated in the following sections.

\section{Resulting Spectral Energy distribution for the YMSC W43}\label{sec:results}

In Figures~\ref{fig:gamma_SED_shell} and \ref{fig:gamma_SED_central}, % and \ref{fig:bohm_SED}, 
we show our computed models for the observed \(\gamma\)-ray spectral energy distribution (SED) of W43 YMSC,  considering different CR injection properties and diffusion regimes.
In each panel %in Figures~\ref{fig:gamma_SED_shell}, \ref{fig:gamma_SED_central}, and \ref{fig:bohm_SED}, 
we indicate the adopted diffusion coefficient profile, the CR acceleration efficiency \(\eta_{\rm CR}\), the spectral index of the injected CR power law, \(\alpha_{\rm CR}\), and the CR cutoff energy.
%(see also Table \ref{tab:bestfit}).
The blue curves represent the predicted \(\gamma\)-ray SED after propagation to Earth. The black dashed curves show the contribution from \(\pi^0\)-decay emission produced 
through the p-p channel, while the red dash-dotted curves represent the inverse Compton scattering (ICS) component. The gray solid curves show the spectrum of CRs that escape from the cluster. All fluxes are normalized to the distance of W43, \(d = 5.5\)~kpc \citep[][]{2024A&A...685A.101L}.

To scale our models to a physical energy injection consistent with observations, we have proceeded as described in Appendix \ref{appendix:luminosity_scaling}.

Before discussing the resulting SEDs, we describe below the normalization adopted for the diffusion coefficient in the Monte Carlo simulations of CR propagation, as well as the parameters assumed for the CR spectrum injected into the three-dimensional domain of the YMSC.

\subsection{CR diffusion normalization}
\label{subsec:initial-cond}

\begin{table}[]
\centering
\begin{tabular}{cccc}
\hline
\textbf{Simulation} & \(\boldsymbol{D_{\rm scale}}\) & $\boldsymbol{\alpha}$ & \textbf{Source} \\ \hline
S1                  & 0.026                & 1/3               & Shell           \\
S2                  & $2.730\times10^{-5}$ & 0.7               & Shell           \\
C1                  & 0.026                & 1/3               & Central         \\
C2                  & $2.730\times10^{-5}$ & 0.7               & Central         \\
B                   & $2.164\times10^{-6}$ & 1                 & Central         \\ \hline
\end{tabular}
\caption{Normalization values of the CR diffusion coefficient used in the Monte Carlo simulations of CR propagation within the YMSC, which are obtained for the different CR-source-diffusion models considered in this work. The labels S1 and S2 denote shell CR-source models with diffusion coefficients $D \propto E^{1/3}$ and $D \propto E^{0.7}$, respectively.
The labels C1 and C2 denote central CR-source models with the same respective diffusion coefficients. The label B denotes the Bohm diffusion model,  with a central source (see text for details).}
\label{tab:input_crpropa}
\end{table}

As stressed, we model CR propagation in the MHD background of the simulated YMSC shown in Figure~\ref{fig:cluster_3D}, using the CRPropa code \citep{2016JCAP...05..038A, 2022JCAP...09..035A}, in particular its low-energy extension~\citep{merten2017a}. 
The diffusion power-law index, $\alpha$, is chosen according to the diffusion model under consideration, as discussed in Section~\ref{sec:cr_diffusion_coeff}. 

In the CRPropa propagator module \texttt{DiffusionSDE}, the diffusion coefficient is defined as:
\begin{align}
    D(R_{\rm CR}) = D_{\rm scale}\cdot 6.1\times10^{24} \cdot \left(\frac{R_{\rm CR}}{4\times10^9V}\right)^\alpha\,\mathrm{m^2 \, s^{-1}}
\end{align}
where \(D_{\rm scale}\) is the input scaling factor that depends on the chosen diffusion model, and $R_{\rm CR}=pc/Ze$ is the CR rigidity, which for particles with energy $E \gg mc^2$, is $R_{\rm CR}=E/Ze$. 
Since we  consider protons only, Z=1 and, for an energy of 1 PeV,  $R_{\rm CR} = 10^{15}$V. We can then calculate $D_{\rm scale}$ comparing the equation above with Equation \ref{eq:diffusion_coefficient} to obtain:

\begin{align}
    D_{\rm scale,\alpha} = \frac{D_{1 {\rm PeV},\alpha}}{6.1\times10^{24}}(4\times10^{-6})^\alpha,
\end{align}
where  $D_{1 {\rm PeV},\alpha}$  is defined in eqs. 2, 3 and 4, for the three diffusion models discussed in this work.

\subsection{$\chi^2$ minimization analysis}\label{sec:chi2_min_analysis}

To quantify the dependence of the predicted SEDs on the CR injection spectrum, we performed a $\chi^2$ analysis varying three key parameters: the spectral power-law index $\alpha_{\rm CR}$, the exponential cutoff energy $E_{\rm cutoff}$, and the CR injection power $P_{\rm CR}$. For all  models, we explored cutoff energies in the range $E_{\rm cutoff} \in [10^{13.5}, 10^{16}]$~eV.

The adopted ranges of $\alpha_{\rm CR}$ and $P_{\rm CR}$ depend on the assumed diffusion regime and source geometry. For the models with $D \propto E^{1/3}$, we used $\alpha_{\rm CR} \in [1.7, 2.2]$. In this case, the CR injection power was varied within $\log_{10}(P_{\rm CR}/\text{erg\,s}^{-1}) \in [37.4, 38.2]$ for the shell-like source geometry, model S1, and within $\log_{10}(P_{\rm CR}/\text{erg\,s}^{-1}) \in [36.8, 37.7]$ for the centrally concentrated geometry, model C1.

For the models with $D \propto E^{0.7}$, namely S2 and C2, we adopted a harder injection spectrum, with $\alpha_{\rm CR} \in [1.4, 1.8]$. The corresponding CR injection power ranges were $\log_{10}(P_{\rm CR}/\text{erg\,s}^{-1}) \in [36.3, 37.1]$ for the shell-like geometry, model S2, and $\log_{10}(P_{\rm CR}/\text{erg\,s}^{-1}) \in [35.7, 36.5]$ for the centrally concentrated geometry, model C2.

Finally, for the Bohm-like diffusion case, $D \propto E$, assuming a central source, model B, we considered $\alpha_{\rm CR} \in [1.0, 1.7]$ and $\log_{10}(P_{\rm CR}/\text{erg\,s}^{-1}) \in [36.4, 37.2]$.

The best-fit parameter values obtained for each model are listed in Table~\ref{tab:bestfit}. We also quantify the CR injection efficiency as
\begin{equation}
\eta_{\rm bol} = \frac{P_{\rm CR}}{L_{\rm W43}},
\end{equation}
where $\eta_{\rm bol}$ measures the fraction of the bolometric luminosity of W43 converted into CR power. Here, we adopt $L_{\rm W43} \sim 10^7 L_\odot$ \citep[][]{2024A&A...685A.101L}. Further details of the $\chi^2$ minimization procedure are provided in Section~\ref{sec:chi2_min_parametric_space}.

\begin{table*}[tb]
    \centering
    \begin{tabular}{cccccc}
        \hline
        \textbf{ID} & $\alpha_{CR}$ & \textbf{$\text{E}_\text{cutoff}$}[eV] & \textbf{$\eta_{bol}=P_{CR}/L_{\text{W43}}$} & \textbf{$P_{CR}$}[erg/s] & \textbf{$\chi^2_{min}$} \\
        \hline
        S1 & $1.95^{+0.03}_{-0.03}$ & $1.40^{+1.01}_{-0.46} \times 10^{14}$ & $1.32\times10^{-3}$ & $5.06^{+1.06}_{-0.40}\times10^{37}$ & 10.84 \\
        S2 & $1.56^{+0.02}_{-0.04}$ & $1.07^{+0.55}_{-0.15} \times 10^{14}$ & $1.06\times10^{-4}$ & $4.07^{+0.76}_{-0.35}\times10^{36}$ & 11.53 \\
        C1 & $1.95^{+0.03}_{-0.03}$ & $1.23^{+0.68}_{-0.56} \times 10^{14}$ & $3.81\times10^{-4}$ & $1.46^{+0.28}_{-0.13}\times10^{37}$ & 7.15  \\
        C2 & $1.58^{+0.02}_{-0.05}$ & $1.24^{+1.18}_{-0.43} \times 10^{14}$ & $2.72\times10^{-5}$ & $1.05^{+0.24}_{-0.07}\times10^{36}$ & 8.84 \\
        B  & $1.35^{+0.02}_{-0.03}$ & $1.57^{+1.14}_{-0.35} \times 10^{14}$ & $1.30\times10^{-4}$ & $4.96^{+0.88}_{-0.43}\times10^{36}$ & 12.37 \\
        \hline
    \end{tabular}
    \caption{CR injection best-fit parameters for the different diffusion coefficients and CR source-location models with their respective \(1\sigma\) variation (see Appendix \ref{appendix:best_fit}). The labels S1 and S2 denote shell CR-source models with diffusion coefficients $D \propto E^{1/3}$ and $D \propto E^{0.7}$, respectively. The labels C1 and C2 denote central CR-source models with the same respective diffusion coefficients. The label B denotes the Bohm diffusion model,  with a central source.
    }\label{tab:bestfit}
\end{table*}

% \bete{[ESTA PARTE ABAIXO QUE DESCREVE O SCALIING VAI PARA UM APENDICE]}

\begin{figure*}[ht]
    \centering
    \includegraphics[width=1.99\columnwidth]{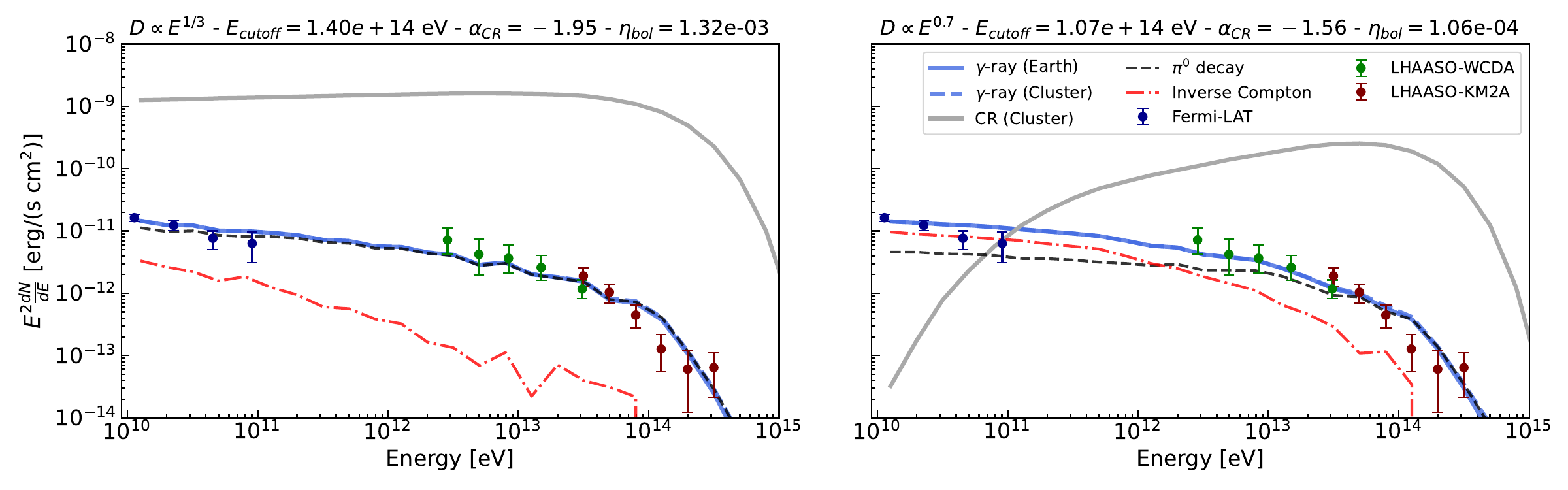}
    \caption{
    Spectral energy distributions (SEDs) of the $\gamma$-ray emission obtained
    at Earth for the spherical-shell CR injection models (S1 and S2 in Table \ref{tab:bestfit}), representing particle
    acceleration at the collective wind shock of the massive stars in the YMSC.
    The observational data correspond to the measured spectrum of W43, with the
    different symbols indicating the Fermi-LAT, LHAASO-WCDA, and LHAASO-KM2A measurements \citep{2025SCPMA..6879502C}, as indicated in the legend.
    The blue solid curves show the total modeled $\gamma$-ray spectrum after
    propagation to Earth, scaled to the distance of W43. The black dashed curves
    show the hadronic contribution produced mainly by proton--proton interactions and
    subsequent $\pi^0$ decay, while the red dash-dotted curves indicate the
    inverse Compton scattering component. The gray solid curves show the
    spectrum of CRs escaping from the cluster.
    The CR injection parameters and the adopted diffusion model are given at the
    top of each panel. The left panel corresponds to the reference
    mirror+scattering diffusion model, with $D(E)\propto E^{1/3}$, whereas
    the right panel shows the comparison case with $D(E)\propto E^{0.7}$,
    associated with a less favored mirror-dominated diffusion regime. 
    }
    \label{fig:gamma_SED_shell}
\end{figure*}

\begin{figure*}
    \centering
    \includegraphics[width=1.99\columnwidth]{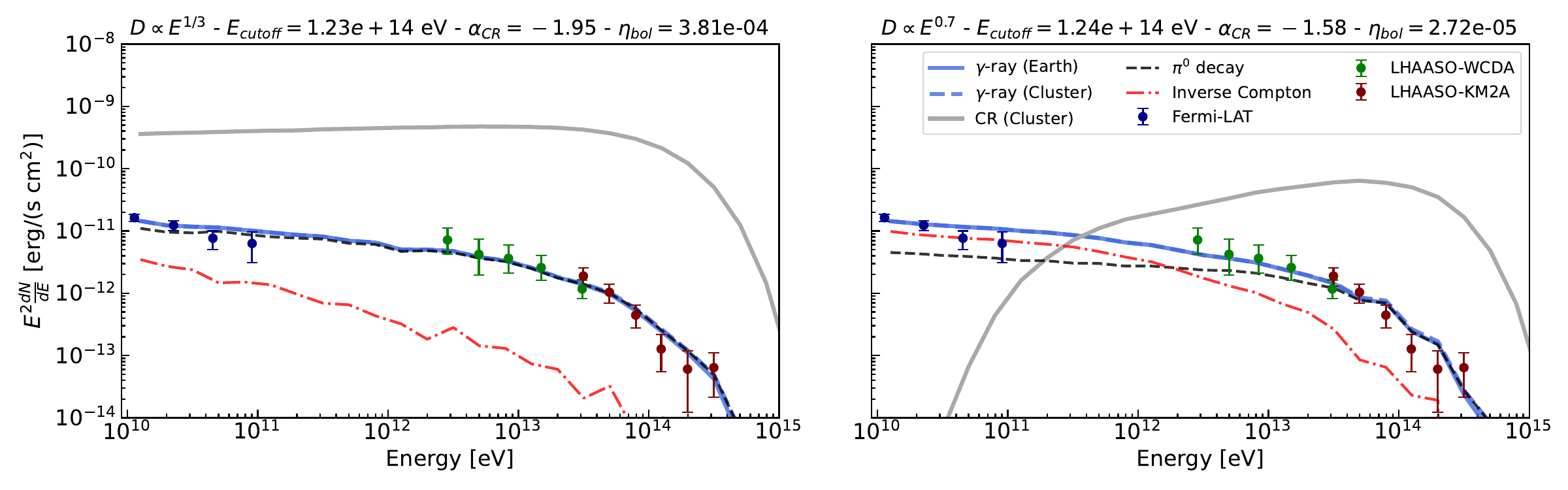}
    \caption{Same as Figure \ref{fig:gamma_SED_shell}, but considering a CR source at the center of the simulated box (models C1 and C2 in Table \ref{tab:bestfit}). While we use the same two regimes for CR diffusion, the total CR power is about one order of magnitude smaller compared to the models presented in Figure \ref{fig:gamma_SED_shell}.
    }
    \label{fig:gamma_SED_central}
\end{figure*}

\subsection{Spherical Shell CR Source Scenario Results}
\label{subsec:shell_results}

Figure~\ref{fig:gamma_SED_shell} shows the resulting $\gamma$-ray spectral
energy distribution (SED) for the scenario in which CRs are injected in a
spherical shell of radius 10~pc. This configuration is intended to mimic CR
acceleration at the collective wind shock produced by the massive stars in the
computed YMSC. The two panels compare different assumptions for the energy dependence of
the diffusion coefficient, illustrating how the diffusion regime affects both
the required CR injection spectrum and the relative importance of the radiative
channels.

In the left panel, we show our reference case, corresponding to
mirror+scattering diffusion, with \(D(E)\propto E^{1/3}\). This model
provides a good fit to the observed W43 spectrum with an injected CR spectrum
close to \(E^{-2}\), as expected for standard Fermi-like acceleration, and a CR
power of order \(\sim 10^{37}\,{\rm erg\,s^{-1}}\). In this case, the
$\gamma$-ray emission is dominated by hadronic interactions, through
proton--proton collisions followed by \(\pi^0\) decay. The inverse Compton
scattering (ICS) component contributes mainly at the lowest energies shown,
near \(10^{10}\)~eV, but remains approximately one order of magnitude below the
\(\pi^0\)-decay component over most of the relevant energy range.

The right panel shows the comparison case with \(D(E)\propto E^{0.7}\),
associated with a less favored, mirror-dominated, diffusion regime. In this
case, the observed spectrum can be reproduced only with a significantly harder
injection spectrum, with an index 
approximately %\(1.5\) and
\(1.6\),
and with a lower CR power, of order \(\sim 10^{36}\,{\rm erg\,s^{-1}}\). The
relative contributions of the emission channels also change: ICS becomes
dominant at lower $\gamma$-ray energies, whereas the \(\pi^0\)-decay component
dominates at higher energies. This behavior reflects the different residence
times and interaction probabilities implied by the stronger energy dependence
of the diffusion coefficient.

The escaping CR spectra further illustrate the effect of the transport model.
For the \(D(E)\propto E^{0.7}\) case, CRs with energies below
\(\sim 10^{11}\)~eV are more efficiently confined and lose a larger fraction
of their energy inside the cluster before escaping. This suppression is less
pronounced in the \(D(E)\propto E^{1/3}\) model. 

Overall, the shell-injection
results show that the observed $\gamma$-ray spectrum of W43 can be reproduced
under different diffusion prescriptions, but the physically preferred
mirror+scattering model naturally requires an injection index close to
\(2\) and produces a predominantly hadronic $\gamma$-ray signal.

\subsection{Central Source Scenario Results}
\label{subsec:central_results}

Figure~\ref{fig:gamma_SED_central} shows the resulting $\gamma$-ray spectrum
for the case in which CRs are injected by a source located at the center of the simulated YMSC. This scenario is intended to represent a compact accelerator
embedded in the cluster, such as a pulsar or a microquasar.

A comparison with the shell-injection case shown in
Figure~\ref{fig:gamma_SED_shell} indicates that, for a fixed diffusion regime,
the final $\gamma$-ray SED is only weakly affected by the assumed source
geometry. The main difference is the normalization required to reproduce the
observed W43 spectrum: the shell model requires a CR injection power about four
to five times larger than the corresponding central-source model. This can be
understood because, in the shell scenario, CRs are injected closer to the outer
regions of the cluster and therefore spend, on average, less time propagating
through the dense central gas before escaping. As a result, a larger injected
CR power is needed to produce the same level of $\gamma$-ray emission.

Despite this difference in normalization, the relative contributions of inverse
Compton scattering and $\pi^0$-decay emission follow similar trends in the two
source geometries. For \(D(E)\propto E^{1/3}\), the emission is predominantly
hadronic, with the $\gamma$-ray flux mainly produced by proton--proton
interactions followed by $\pi^0$ decay. For \(D(E)\propto E^{0.7}\), inverse
Compton scattering contributes more significantly at lower energies, while the
$\pi^0$-decay component dominates at higher energies.

The main spectral difference appears in the escaping CR component. In the
central-source case, CRs must cross a larger fraction of the dense cluster
environment before leaving the system, which increases their probability of
interacting inside the cluster, especially in the range
\(10^{10}\)--\(10^{11}\)~eV. Consequently, the central-source models require
lower CR injection powers, by a factor of approximately four to five, than the
corresponding shell models to reproduce the observed W43 spectrum. Overall,
these results indicate that the diffusion regime has a stronger impact on the
shape of the $\gamma$-ray spectrum than the precise location of the CR source
inside the cluster.

\subsection{$\chi-$Square Minimization of the Free Parametric Space}\label{sec:chi2_min_parametric_space}

In Figures \ref{fig:gamma_SED_shell}  and \ref{fig:gamma_SED_central}, we have presented our best computed models for the W43 YMSC, obtained by combining the 3D MHD background model with Monte Carlo CR propagation and radiative cascading calculations. These models were computed using the  three CR free parameters, initially fixed at the best-fit values obtained from $\chi^2$ analysis and listed in Table \ref{tab:bestfit}.

We now show the \(\chi^2\) minimization procedure around this fiducial parameter set in order to evaluate how sensitive our results are to variations in these free parameters.

\begin{figure*}[ht]
    \centering
    \includegraphics[width=1.99\columnwidth]{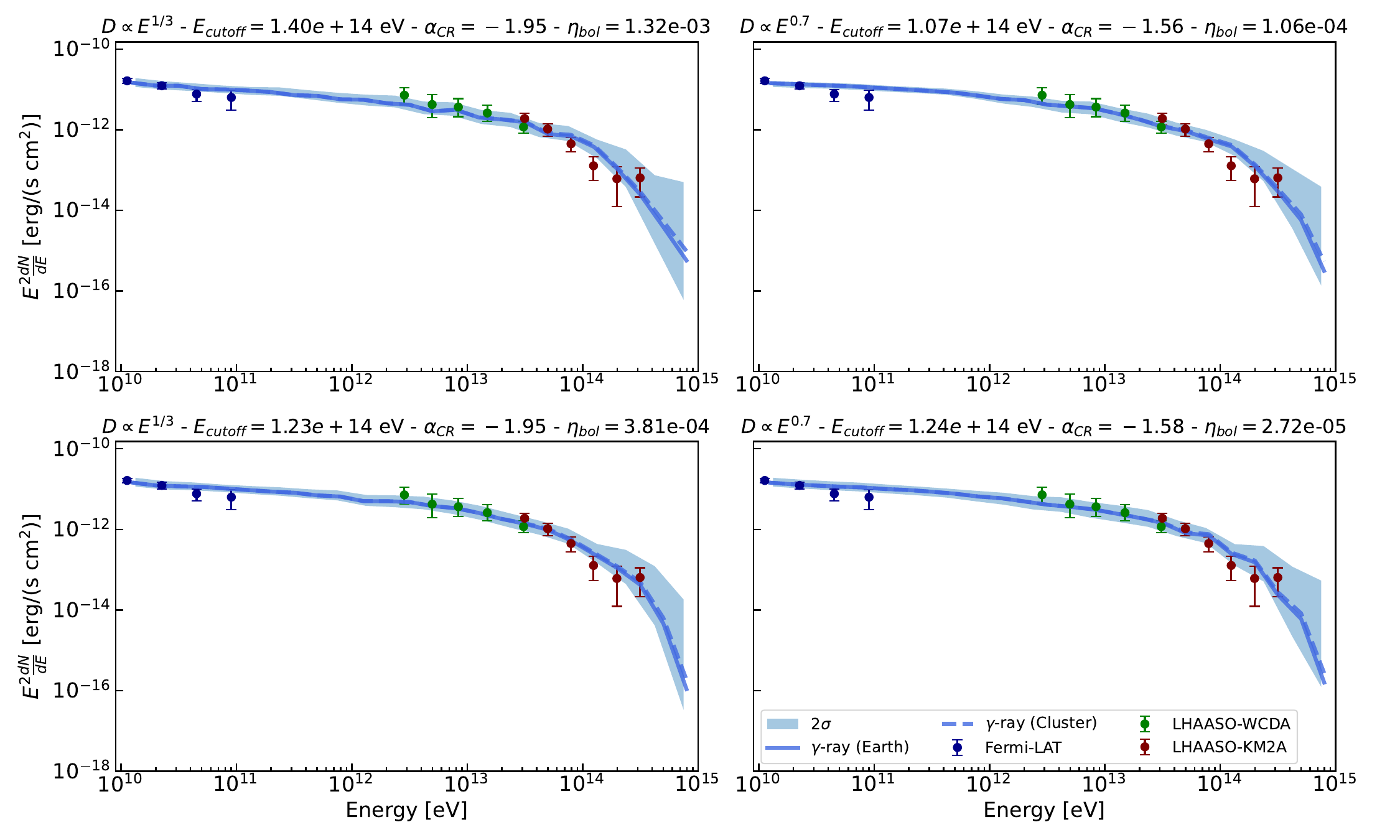}
    \caption{SED variation considering the $2\sigma$ parameters in Figure \ref{fig:chi2_maps}. Light-blue line and dashed-light-blue line show the best fit SEDs for each model (same as in Figures \ref{fig:gamma_SED_shell} and \ref{fig:gamma_SED_central}). Shaded region shows the SED variation considering the \(2\sigma\)  parameter region shown in Figure \ref{fig:chi2_maps}. Dark-blue, green and 
    %dark-red 
    brown dots show the observed SED of W43.}
    \label{fig:gamma_SED_chi2_variation}
\end{figure*}

The last column of Table~\ref{tab:bestfit} gives the minimum $\chi^2$ value associated with each best-fit parameter set. Figure~\ref{fig:gamma_SED_chi2_variation} quantifies the dependence of our results on the adopted diffusion prescription and source geometry by showing the range of SEDs obtained within the allowed parameter space around the best-fit solutions.

The mirror+scattering models show the best statistical agreement when compared to the mirror dominant regime. Also, the mirror+scattering models require an injected spectrum close to \(E^{-2}\), with \(\alpha_{\rm CR}\simeq 1.96\), reinforcing the physical preference for the \(D(E)\propto E^{1/3}\) diffusion regime. 

% The best statistical
% agreement is obtained for the central-source mirror+scattering model, with
% \(\chi^2_{\rm min}=7.15\), followed by the shell mirror+scattering case, with
% \(\chi^2_{\rm min}=10.379\). Both models require an injected spectrum close to
% \(E^{-2}\), with \(\alpha_{\rm CR}\simeq 1.96\), reinforcing the physical
% preference for the \(D(E)\propto E^{1/3}\) diffusion regime. 

The comparison
models with \(D(E)\propto E^{0.7}\) also provide acceptable fits, but only with
harder injection spectra, \(\alpha_{\rm CR}\simeq 1.56-1.59\). This again
indicates that, although different diffusion prescriptions can reproduce the
observed W43 SED, the mirror+scattering model gives the most natural solution
from the standpoint of CR propagation theory. The table also shows that the
required CR power is systematically lower for the central-source geometry than
for the shell geometry, consistent with the longer residence time and higher
interaction probability of particles injected near the cluster center.

\subsection{Comparison with Bohm Diffusion}

\begin{figure}
    \centering
    \includegraphics[width=1.\columnwidth]{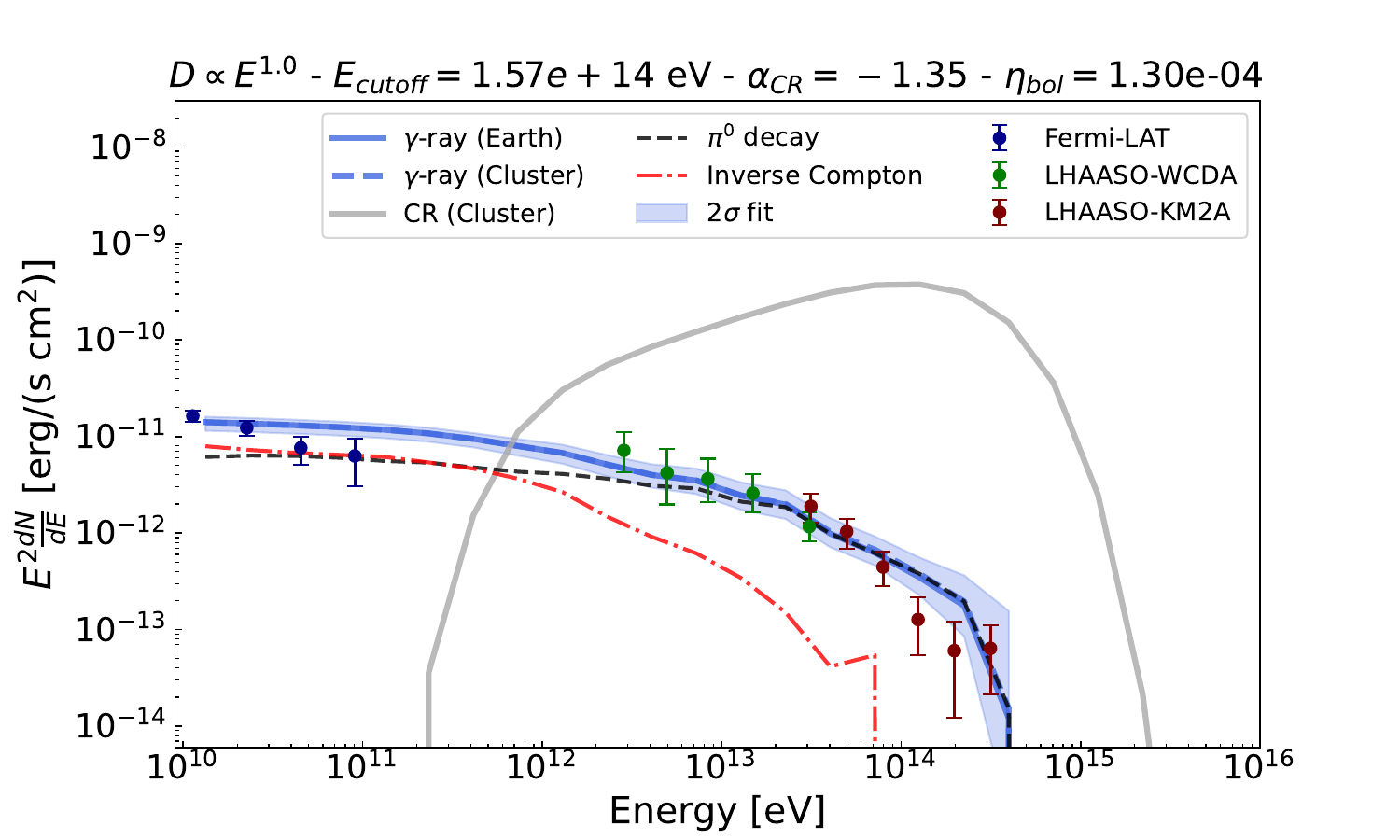}
    \caption{Same as Figures \ref{fig:gamma_SED_shell} and \ref{fig:gamma_SED_central}, but considering $D\propto E^{1.0}$ (Bohm diffusion). For this model (B1), only a central source was considered. The shaded region shows, as in Figure \ref{fig:gamma_SED_chi2_variation}, the SED variation considering the $2\sigma$ parameter region shown in Figure \ref{fig:chi2_maps}.}
    \label{fig:bohm_SED}
\end{figure}

Finally, we have also tested the scenario where $D\propto E^{1.0}$ (Equation \ref{eq:diffusion_coefficient_bohm}), usually referred as Bohm diffusion in the literature, and considering a central source of CRs. The resulting spectra is shown in Figure \ref{fig:bohm_SED} (similar to Figures \ref{fig:gamma_SED_shell}--\ref{fig:gamma_SED_chi2_variation}). The resulting $\gamma$'s are also a mixture of IC process and p-p  $\pi^0$ decay, as in Figure \ref{fig:gamma_SED_central}, and similar to what we see when $D\propto E^{0.7}$. However, the shape of the spectrum does not visually follow the low energy trend and most of the CRs are absorbed inside source, with only CRs with energies above $10^{11}$~eV surviving. Moreover, the CR injection required to reproduce the observation needs to be even harder, with \(\alpha = -1.37\).

\section{Discussion}
\label{sec:discussion}
The results presented above show that the observed $\gamma$-ray spectrum of
W43 can be reproduced under different assumptions for the CR diffusion
coefficient and source geometry. However, the physical interpretation of the
best-fit models depends strongly on the adopted transport prescription.

We first consider the shell-injection scenario, which mimics CR acceleration at
the collective wind termination shock produced by the massive stars in the
cluster. For the fiducial mirror+scattering diffusion model,
\(D(E)\propto E^{1/3}\), the observed W43 spectrum is reproduced with an
injected CR spectrum close to \(E^{-2}\). This is consistent with the spectral
index expected for Fermi-like acceleration and therefore provides a natural
connection between the acceleration mechanism and the observed VHE emission. In
this case, the $\gamma$-ray emission is dominated by the hadronic component,
mainly through proton--proton interactions followed by \(\pi^0\) decay, while
inverse Compton scattering contributes mostly at the lower-energy end of the
spectrum.
The comparison model with \(D(E)\propto E^{0.7}\), which describes a less realistic scenario where mirror diffusion is dominant over scaterring,  can also fit the data, but it
requires a significantly harder injected CR spectrum, with an index close to
\(1.6\). Such a hard spectrum is less natural for standard diffusive shock
acceleration and therefore makes this transport prescription less physically
favored for W43. In this case, inverse Compton scattering becomes more important
at lower $\gamma$-ray energies, whereas the \(\pi^0\)-decay component dominates
at higher energies. This demonstrates that the relative importance of leptonic
and hadronic emission channels is strongly controlled by the assumed diffusion
regime.

The central-CR-source models lead to very similar $\gamma$-ray spectral shapes for
a given diffusion coefficient. The main difference with respect to the shell
scenario is the required CR injection power. Since particles injected at the
cluster center propagate through a larger fraction of the dense gas before
escaping, they have a higher probability of interacting inside the cluster.
Consequently, the central-source models require a lower CR power than the
corresponding shell-injection models to reproduce the observed W43 flux. This
indicates that the $\gamma$-ray spectrum alone cannot unambiguously distinguish
between a compact central accelerator, such as a pulsar or microquasar, and
acceleration at the collective wind shock of the massive stars.

Our results are broadly consistent with previous studies of W43. For example,
\cite{2025SCPMA..6879502C} derived a CR injection index of approximately
\(2.3\) by fitting hadronic interaction models to the measured $\gamma$-ray
spectrum with {\sc naima} package, assuming that the observed emission traces the parent
proton population. 
\cite{2025icrc.confE.692I} reports a somewhat harder index, closer to our mirror+scattering model, with an injection
spectrum close to \(E^{-2}\). However they consider a larger region for the wind termination shock.
The difference arises because our calculation
includes CR propagation in a turbulent magnetized YMSC environment, rather than
inferring the parent proton spectrum directly from the observed $\gamma$-ray
emission.
A key point of this work is that the \(D(E)\propto E^{1/3}\) dependence is not
introduced merely as a phenomenological fitting choice. Instead, it is motivated
by CR transport theory in turbulent magnetized media, where magnetic mirroring
and pitch-angle scattering act together to suppress diffusion. In the YMSC
environment considered here, the turbulence is mostly supersonic and
sub-Alfv\'enic, mainly in the inner parts of the cluster, implying that the magnetic field is dynamically important.
Under these conditions, fast MHD modes can efficiently scatter CRs, while
spatial variations in the magnetic-field strength generate magnetic mirrors
that increase the particle residence time. The combination of these effects
naturally leads to a suppressed diffusion coefficient with a weak energy
dependence, making the mirror+scattering model the most physically motivated
case among those explored here.

We have also considered the effect of $\gamma$-ray absorption along the line of
sight toward W43. The simulations indicate that absorption is weak across the
energy range considered. The largest attenuation occurs around
\(\sim 10^{15}\,{\rm eV}\) mainly because of CMB photons, where the spectrum is reduced by at most a factor of
about two. Given the location of W43 in the Galaxy, gamma-ray absorption by ISRF photons above tens of TeV is nearly negligible. Therefore, absorption in the Galactic environment does not significantly modify the inferred CR
properties or the main conclusions regarding the preferred diffusion regime.

Finally, the Bohm-like case, \(D(E)\propto E\), was included as a limiting
comparison. Although this prescription is often used to represent strong
scattering, it does not capture the combined effects of mirror trapping and
fast-mode scattering expected in the turbulent YMSC environment. Moreover, the
resulting spectrum provides a less satisfactory description of the low-energy
part of the observed emission and requires an unrealistically low CR injection
power compared with the other models. We therefore regard the Bohm-like case as
a useful limiting prescription, but not as the preferred description of CR
transport in W43. 

\section{Conclusions}\label{sec:conclusion}
We have investigated CR propagation and $\gamma$-ray production in young massive
stellar clusters by combining a 3D MHD simulation of a turbulent magnetized
cluster environment with Monte Carlo CR propagation and cascading calculations
(performed with {\sc CRPropa}). We applied this framework to the massive
star-forming region W43 and compared the resulting spectra directly with the
observed $\gamma$-ray emission measured by Fermi-LAT and LHAASO.
Our model includes the 3D distributions of gas density and magnetic field from
the MHD simulation, together with the relevant photon fields inside and around
the cluster, including stellar radiation,  the Galactic
interstellar radiation field, and the cosmic microwave background. We considered
two possible CR injection geometries: a compact source located at the center of
the cluster, and a spherical shell representing particle acceleration at the
collective wind shock of the massive stars.
The main result of this work is that the W43 $\gamma$-ray spectrum is naturally
reproduced when CR transport is described by the mirror+scattering diffusion
model, with
\begin{equation}
    D(E)\propto E^{1/3}.
\end{equation}
This diffusion law is physically motivated by CR transport in turbulent
magnetized media, where magnetic mirroring and pitch-angle scattering act
together to suppress particle escape. Unlike purely phenomenological
prescriptions, this model follows from the expected properties of MHD turbulence
in the YMSC environment.

For the fiducial mirror+scattering model, the observed spectrum is reproduced
with an injected CR spectrum close to \(E^{-2}\), as expected for Fermi-like
acceleration. The best-fit CR injection powers are
\(\sim 1.46\times10^{37}\,{\rm erg\,s^{-1}}\) for the central-source model and
\(\sim 5.06\times10^{37}\,{\rm erg\,s^{-1}}\) for the shell-injection model.
The larger power required in the shell case reflects the shorter residence time
of particles injected closer to the outer regions of the cluster.
We find that the dominant $\gamma$-ray production channel in the fiducial model
is hadronic emission from proton--proton interactions followed by \(\pi^0\)
decay. Inverse Compton scattering contributes mainly at lower $\gamma$-ray
energies, but remains subdominant over most of the observed range. 

Alternative diffusion prescriptions, such as \(D(E)\propto E^{0.7}\) where only mirror diffusion prevails, can also reproduce
the data, but require a much harder CR injection spectrum, with index
\(\alpha_\text{CR} \sim 1.6\), and are therefore less favored from the standpoint of standard
acceleration theory. The Bohm-like case provides only a limiting comparison and
does not offer as natural a description of the W43 spectrum.

We also find that the final $\gamma$-ray spectral shape is more sensitive to the
diffusion regime than to the precise location of the CR source. Central and
shell-like injection geometries produce similar spectra for a given diffusion
coefficient, although they require different CR injection powers. Therefore,
the current $\gamma$-ray spectrum alone cannot distinguish whether the dominant
accelerator is a compact central source, such as a pulsar or microquasar, or
the collective wind shock associated with the massive-star population.
Multiwavelength and multi-messenger observations, together with forthcoming high-angular-resolution $\gamma$-ray facilities, will be essential to break this degeneracy. In particular, X-ray and optical constraints may help identify possible compact accelerators, while forthcoming facilities such as the ASTRI Mini-Array \citep{2022JHEAp..35....1V} and CTAO \citep{2019scta.book.....C, 2023arXiv230512888H} will provide improved morphological information on the embedded $\gamma$-ray sources. Neutrino observations offer a complementary probe of the hadronic scenario; in Appendix~\ref{appendix:neutrinos_results} we present preliminary results for the predicted neutrino flux from our W43 models. Such multiwavelength and multi-messenger observations will be crucial for determining whether the emission is centrally concentrated or instead associated with an extended wind-driven shell.

Overall, our results support a scenario in which W43 acts as an efficient
cosmic-ray accelerator embedded in a turbulent magnetized environment. The
agreement between the observed spectrum and the mirror+scattering diffusion
model suggests that MHD turbulence plays a central role in regulating CR
transport in young massive stellar clusters and in shaping their VHE
$\gamma$-ray emission.

\section*{Acknowledgements}
L.B-M. acknowledges support from the Brazilian Funding Agencies CAPES, CNPq and FAPESP; E.M.G.D.P. also acknowledges support from FAPESP (grants 2013/10559-5, 2021/02120-0) and CNPq (grant 308643/2017-8).
The simulations presented in this work were performed using the cluster of the Group of Plasmas and High-Energy Astrophysics at IAG-USP (GAPAE), acquired with support from FAPESP (grants 2013/10559-5 and 2021/02120-0). R.A.B. is supported by the Agence Nationale de la Recherche (ANR), project ANR-23-CPJ1-0103-01. G.D.M.'s work is supported by \textit{FPI Severo Ochoa} PRE2022-101820 grant and also by the grants PID2024-155874NB-C21 and CEX2020-001007-S (before PID2021-125331NB-I00 and CEX2020-001007-S), both funded by MCIN/AEI/10.13039/501100011033 and by ``ERDF A way of making Europe''. G.D.M. also acknowledges the MultiDark Network, ref. RED2022-134411-T.; Also, special thanks to Yaocheng Cheng for the helpful corrections to the manuscript.

\appendix

\section{Photon field calculation}\label{appendix:photon_field}

\subsection{Blackbody radiation from the stars}\label{appendix:star_blackbody}

We estimate the radius and effective temperature of each star from its mass using the approximate scaling relations \citep[e.g.][]{2009itss.book.....P,2013sse..book.....K}
\begin{align}
    R_i &= R_{\odot}\left(\frac{M_i}{M_{\odot}}\right)^{0.8},\\
    T_i &= 5778\,{\rm K}\left(\frac{M_i}{M_{\odot}}\right)^{0.5}.
\end{align}

The monochromatic luminosity of each star is then given by
\begin{align}
    L_i(\lambda,T_i) = 4\pi R_i^2\,\pi B_{\lambda}(T_i,\lambda),
\end{align}
where $B_{\lambda}(T_i,\lambda)$ is the Planck function. Assuming that the stellar radiation is isotropically distributed within the cluster of radius $R_{\rm cluster}$, the mean radiation energy density per unit wavelength contributed by star $i$ can be written as
\begin{align}
    u_{{\rm avg},i}^{\rm tot}(\lambda) =
    \frac{3L_i(\lambda,T_i)}{4\pi c R_{\rm cluster}^2}.
\end{align}

The total stellar radiation energy density per unit wavelength is obtained by summing over all stars in the simulation snapshot,
\begin{align}
    u_{\rm avg}^{\rm tot}(\lambda) =
    \sum_i u_{{\rm avg},i}^{\rm tot}(\lambda).
\end{align}

Here, 
%$L_i$ is the luminosity per unit wavelength
 $u_{{\rm avg},i}^{\rm tot}(\lambda)$ is the contribution of star $i$ to the volume-averaged radiation energy density inside the cluster. In this treatment, we approximate the stellar radiation field as isotropic; the implications and limitations of this assumption are discussed in Section \ref{sec:interaction_rates}.

\subsection{Bremsstrahlung radiation}\label{appendix:bremsstrahlung_radiation}

To estimate the free-free emission from the hot background gas inside the YMSC, we compute the specific free-free emissivity, $j_\nu^{\rm ff}$, in each cell of the MHD simulation. Following \citet{1986rpa..book.....R}, this emissivity is given by
\begin{equation}
\begin{split}
j_\nu^{\rm ff} &= 5.44 \times 10^{-39}\, Z^2\, n_e\, n_i\, T^{-1/2}\,
\bar{g}_{\rm ff} \\
&\quad \times \exp\left(-\frac{h\nu}{k_B T}\right)
\quad
[\mathrm{erg\,cm^{-3}\,s^{-1}\,Hz^{-1}\,sr^{-1}}],
\end{split}
\end{equation}
where $Z$ is the atomic number, $n_e$ and $n_i$ are the electron and ion number densities, respectively, $T$ is the gas temperature, and $\bar{g}_{\rm ff}$ is the velocity-averaged Gaunt factor. We approximate the latter as
\begin{align}
\bar{g}_{\rm ff} \simeq
\ln\left[
\exp\left(
5.960 - \frac{\sqrt{3}}{\pi}
\ln\left[
\frac{Z\nu_{\rm GHz}}{10(T/10^4\,{\rm K})^{3/2}}
\right]
\right)
+ e
\right].
\end{align}

In the YMSC simulation, we use the FLASH variable \texttt{ihp}, which gives the fraction of the gas density in the form of ionized hydrogen in each cell. We assume $n_e=n_i$ and include only cells with temperatures above $10^4\,{\rm K}$. From the resulting free-free emissivity, we estimate the corresponding photon density field, shown in Figure~\ref{fig:all_photon_fields}.

\subsection{Dust emission}\label{appendix:dust_radiation}

We estimate the spectral energy density of the radiation field produced by
thermal dust emission within the simulated cluster volume as 
modified 
blackbody emitter. Below we present this approximation in more detail.
 
\subsubsection{Modified blackbody emissivity}

Dust grains are treated as modified blackbody (greybody) emitters \citep{1983QJRAS..24..267H,1984ApJ...285...89D}. The
volume emissivity of a cell $i$ at frequency $\nu$ is
 
\begin{equation}
    j_\nu^{(i)} = 4\pi\, \kappa_\nu\, \rho_\mathrm{dust}^{(i)}\, B_\nu\!\left(T_d^{(i)}\right),
    \label{eq:emissivity}
\end{equation}
 
\noindent where $B_\nu(T_d)$ is the Planck function evaluated at the local
dust temperature $T_d^{(i)}$, and $\kappa_\nu$ is the dust opacity modelled
as a power law in frequency,
 
\begin{equation}
    \kappa_\nu = \kappa_0 \left(\frac{\nu}{\nu_0}\right)^{\!\beta},
    \label{eq:opacity}
\end{equation}
 
\noindent with reference opacity $\kappa_0 = 0.1\ \mathrm{m^2\,kg^{-1}}$ at
$\nu_0 = 1\ \mathrm{THz}$ \citep{1990AJ.....99..924B} and emissivity spectral index
$\beta$.  The dust mass density in each cell is obtained from the gas density
and the local dust-to-gas mass fraction $f_\mathrm{dg}$,
 
\begin{equation}
    \rho_\mathrm{dust}^{(i)} = f_\mathrm{dg}\, \rho_\mathrm{gas}^{(i)}.
\end{equation}
 
\subsubsection{Total emitted power and photon energy density}
 
The total emitted power spectrum of the simulation box is obtained by
summing the cell emissivities over all $N = 128^3$ cells, each of volume
$\Delta V = (L/128)^3$,
 
\begin{equation}
    P_\nu^\mathrm{tot} = \sum_{i=1}^{N} j_\nu^{(i)}\,\Delta V
                       = 4\pi\,\kappa_\nu \sum_{i=1}^{N}
                         \rho_\mathrm{dust}^{(i)}\,
                         B_\nu\!\left(T_d^{(i)}\right) \Delta V,
    \label{eq:Ptot}
\end{equation}
 
\noindent where $L$ is the domain (or box) side length of the denser region where dust should contribute the most. For our approximation we adopt $L=20$ pc.  Assuming the
box is optically thin to its own far-infrared emission, each photon travels
freely until it escapes.  The volume-averaged photon energy density is then
 
\begin{equation}
    u_\nu = \frac{P_\nu^\mathrm{tot}}{V_\mathrm{box}\,c} \cdot \frac{L}{2},
    \label{eq:u_nu}
\end{equation}
 
\noindent where $V_\mathrm{box} = L^3$ and the factor $L/2$ is the
characteristic mean path length a photon travels before escaping a
homogeneously emitting cube of side $L$, giving a residence time
$\tau_\mathrm{res} = L/(2c)$.
 
\subsubsection{Spectral energy density}% $E^2\,\mathrm{d}N/\mathrm{d}E$}
 
Converting from frequency to photon energy via $E = h\nu$ and using
$u_\nu\,\mathrm{d}\nu = u_E\,\mathrm{d}E$, the photon number density
spectrum is
 
\begin{equation}
    \frac{\mathrm{d}N}{\mathrm{d}E} = \frac{u_\nu}{h\,E}.
\end{equation}
 
\noindent Multiplying by $E^2$ yields the standard SED representation used
in cosmic-ray physics,
 
\begin{equation}
    E^2\frac{\mathrm{d}N}{\mathrm{d}E} = \nu\,u_\nu,
    \label{eq:E2dNdE}
\end{equation}
 
\noindent which has units of $\mathrm{eV\,m^{-3}}$ after converting from SI.
The result is a volume-averaged quantity representing the mean radiation
field experienced by a cosmic ray at an arbitrary location within the box. The final spectrum is showed in Figure \ref{fig:all_photon_fields}.

\section{Synchrotron losses}\label{appendix:sync_loss}

\begin{figure}
    \centering
    \includegraphics[width=1.0\linewidth]{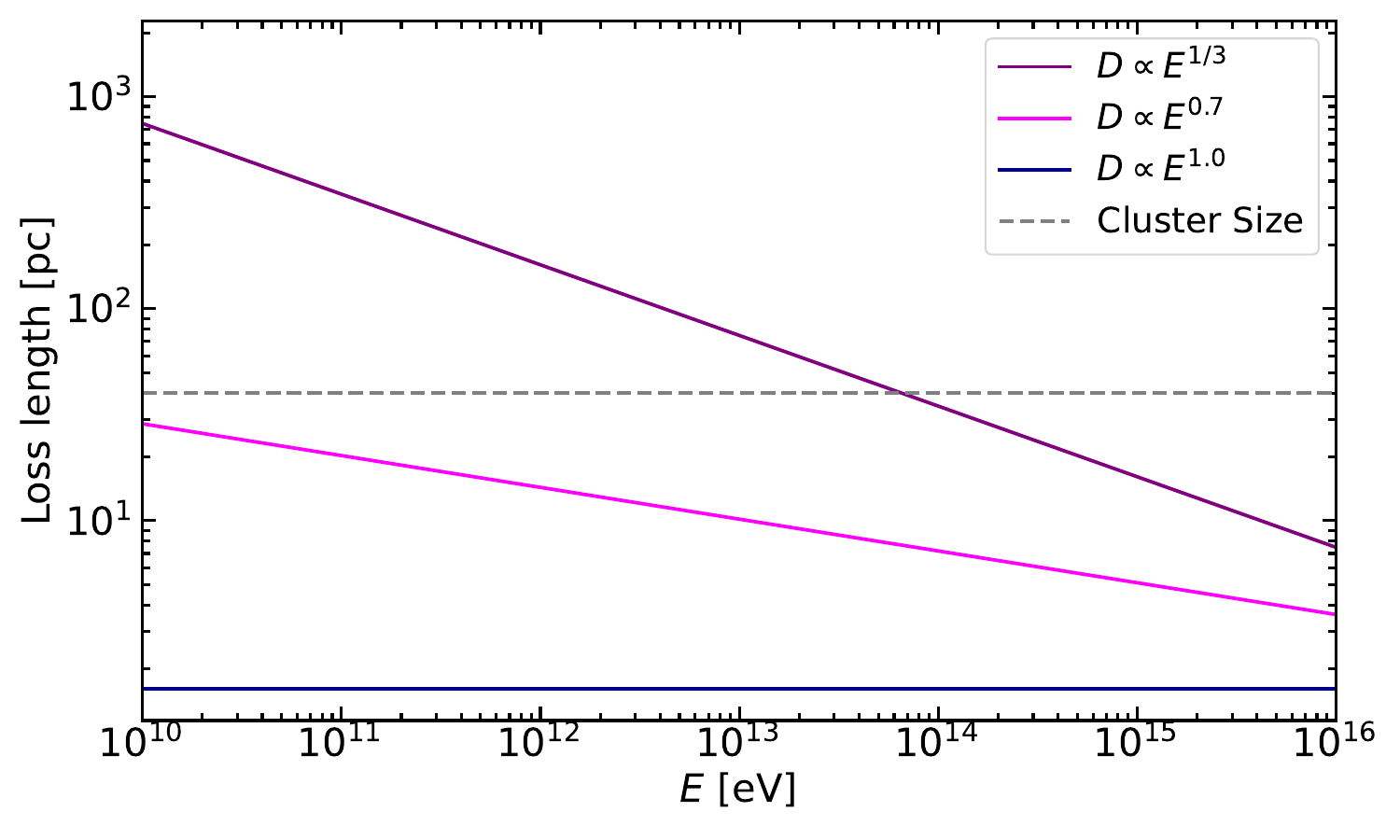}
    \caption{Synchrotron loss length (Equation \ref{eq:loss_length}) as a function of electron energy for three diffusion coefficient scalings: mirror+scattering ($D \propto E^{1/3}$, purple), $D \propto E^{0.7}$ (magenta), and Bohm-like diffusion ($D \propto E^{1.0}$, navy). The dashed gray line marks the simulated cluster size. Curves above the dashed line indicate energies at which electrons can diffuse beyond the cluster before cooling substantially; curves below it indicate energies at which electrons remain confined and radiate their energy in situ.}
    \label{fig:sync_loss}
\end{figure}

The synchrotron cooling time of an electron of energy $E$ in a magnetic field $B$ is
\begin{equation}
    t_{\rm cool}(E,B) = \frac{6\pi m_e c}{\sigma_T B^2 \gamma},
\end{equation}
with $\gamma = E/m_e c^2$ and $\sigma_T$ the Thomson cross section. Combining this with the diffusion coefficient $D(E)$ (Section \ref{sec:cr_diffusion_coeff}) yields the diffusive loss length,
\begin{equation}
    L_{\rm loss}(E) = \sqrt{D(E)\, t_{\rm cool}(E,B)},
    \label{eq:loss_length}
\end{equation}
the characteristic distance an electron travels while diffusing before radiating away an order-unity fraction of its energy. We evaluate Equation~\ref{eq:loss_length} using the region-averaged $B$ field, for the three diffusion regimes discussed in this work.

Figure~\ref{fig:sync_loss} shows $L_{\rm loss}(E)$ for the three regimes, compared to the cluster size (dashed line). In all cases, synchrotron losses are efficient enough to keep $L_{\rm loss}$ within a few hundred parsecs across the full 10~GeV--10~PeV range. For the mirror+scattering case ($D\propto E^{1/3}$), $L_{\rm loss}$ exceeds the cluster size below $\sim7\times10^{13}$~eV, so electrons injected at these energies can escape the cluster before cooling and potentially affect the surrounding medium; above this energy they remain confined and cool in situ. For $D\propto E^{0.7}$ and $D\propto E^{1.0}$, $L_{\rm loss}$ stays below the cluster size across the entire energy range, so electrons remain confined and radiate locally regardless of the energy.

\section{Scaling our models to the observations}\label{appendix:luminosity_scaling}

To scale our CR Monte Carlo simulations to a physically meaningful energy injection rate consistent with observations, we normalize the simulated gamma-ray spectrum at $10\,\mathrm{GeV}$ to the corresponding Fermi-LAT flux. We define the scaling constant as
\begin{align}
    k_{\text{scale}} = \frac{L_{\text{target}}}{\left.E^2 \frac{dN}{dE}\right|_{\text{sim}}(10\,\mathrm{GeV})},
\end{align}
where  $N$ is the  number of gamma-ray photons with energy E per volume,   
$E^2 \frac{dN}{dE}|_{\text{sim}}$
is the simulated gamma-ray spectrum and $L_{\text{target}}$ is a reference power chosen such that
\begin{align}
    \left.E^2 \frac{dN}{dE}\right|_{\mathrm{Fermi}}(10\,\mathrm{GeV}) \approx k_{\text{scale}} \cdot \left.E^2 \frac{dN}{dE}\right|_{\text{sim}}(10\,\mathrm{GeV}).
\end{align}

%As a consistency check on
Using $k_{\text{scale}}$, we compute the corresponding total power injected into cosmic rays,
\begin{align}
    P_{\text{CR}} = k_{\text{scale}} \cdot \sum_i E_{i}\, w_i \quad (\text{erg/s}),
\end{align}
where $E_{i}$ is the energy of the $i$-th injected CR bin and $w_i$ is its weight, set by the injection power-law index $\alpha_{\text{CR}}$.
This value corresponds to a fraction $\eta_\text{CR} < 1$ of the bolometric luminosity of the YMSC system, which is given in Table \ref{tab:bestfit} along with $P_\text{CR}$ for the different models discussed in this work.

\section{Best fit parameter search}\label{appendix:best_fit}

\begin{figure*}%[!ht]
    \centering
    \includegraphics[width=1.98\columnwidth]{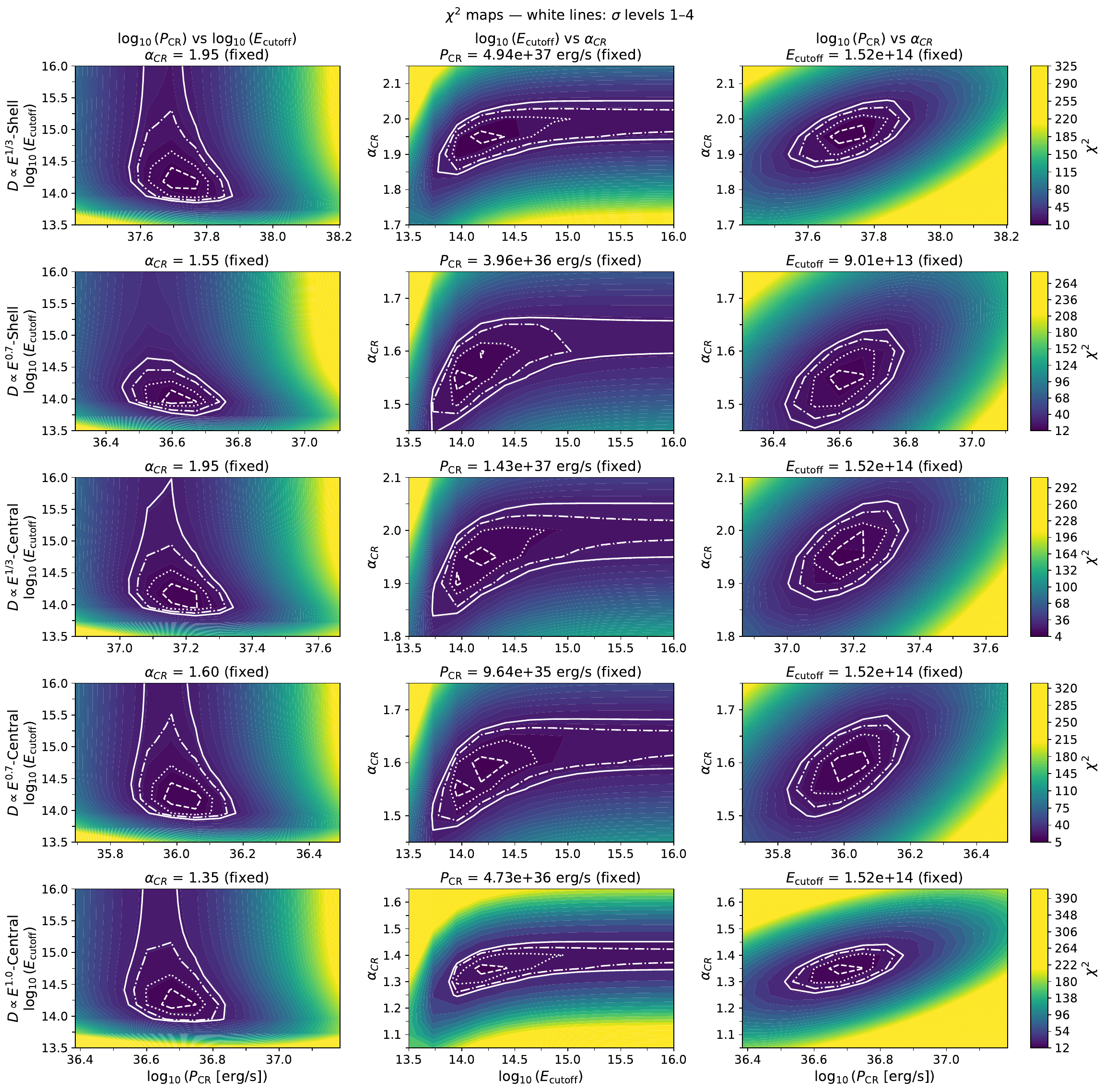}
    \caption{Comparison of $\chi^2$ maps for different fixed CR injection parameters. The left-hand side column shows the maps for fixed CR spectral power law index  parameter ($\alpha_\text{CR}$); the center column shows maps with  fixed CR power  $P_\text{CR}$, and the right-hand side shows maps  for  fixed spectral energy cutoff ($E_\text{cutoff}$). The color scaling is shown in the right-hand side. White lines show the $\sigma=1$ to $\sigma=4$ levels around the best fit value. The corresponding model of each row is shown in the left-hand side of the figure.
    }
    \label{fig:chi2_maps}
\end{figure*}

Figure \ref{fig:chi2_maps} presents the $\chi^2$ maps resulting from the comparison between the model-predicted SEDs and the observed SED of W43, highlighting the dependence of the fit quality on the cosmic-ray injection parameters (see Section \ref{sec:results}).

%\section{Neutrino production: Preliminary 
\section{Predicted Neutrino production }
\label{appendix:neutrinos_results}

\begin{figure*}
    \centering
    \includegraphics[width=1.98\columnwidth]{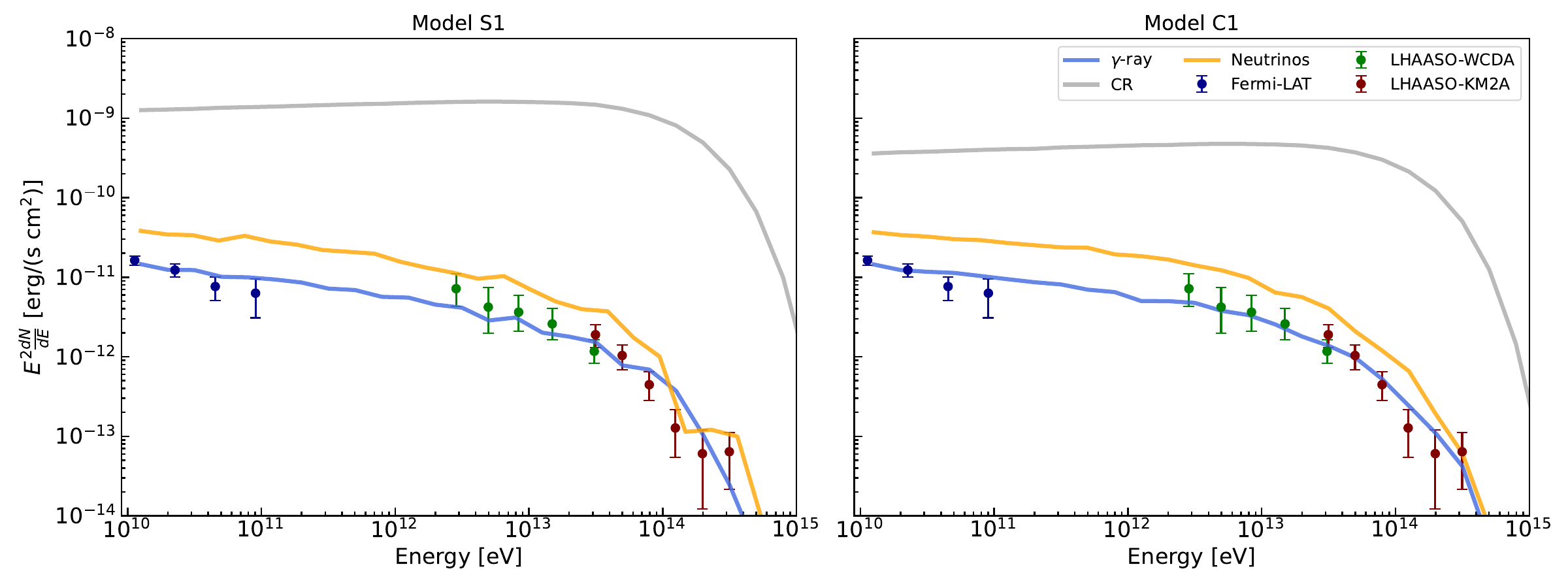}
    \caption{Same as the left panel of Figures~\ref{fig:gamma_SED_shell} and \ref{fig:gamma_SED_central}, now including the predicted neutrino spectra (yellow curve). Left: model S1 (shell injection, $D\propto E^{1/3}$). Right: model C1 (central injection, $D\propto E^{1/3}$).}
    \label{fig:neutrinos}
\end{figure*}

As a by-product of our simulations, we also compute the secondary neutrino production. In Figure~\ref{fig:neutrinos} we show the resulting neutrino flux at Earth for our two fiducial models, S1 and C1 (see Table~\ref{tab:bestfit}). The neutrino spectrum is expected to be a few times more intense than the $\gamma$-ray one and to extend to slightly higher energies, as a natural consequence of the pion production and decay kinematics in $pp$ interactions. While young massive stellar clusters and other starburst regions could plausibly account for a fraction of the diffuse astrophysical neutrino flux observed by IceCube, the detection of an individual source remains challenging for current-generation neutrino telescopes \citep[see, e.g.,][]{2015MNRAS.453..113B,2025icrc.confE.598C}. A detailed assessment of the detection prospects for W43, however, lies beyond the scope of this work.

\bibliographystyle{elsarticle-harv} 
\bibliography{example}



%% For example:

%% \bibitem[Aladro et al.(2015)]{Aladro15} Aladro, R., Martín, S., Riquelme, D., et al. 2015, \aas, 579, A101

%%\end{thebibliography}

\end{document}